\documentclass[lettersize,journal]{IEEEtran}

\usepackage{tikz,xcolor,hyperref}

\definecolor{lime}{HTML}{A6CE39}
\DeclareRobustCommand{\orcidicon}{
\begin{tikzpicture}
\draw[lime, fill=lime] (0,0)
circle[radius=0.16]
node[white]{{\fontfamily{qag}\selectfont \tiny \.{I}D}}; 
\end{tikzpicture}
\hspace{-2mm}
}
\foreach \x in {A, ..., Z}{%
\expandafter\xdef\csname orcid\x\endcsname{\noexpand\href{https://orcid.org/\csname orcidauthor\x\endcsname}{\noexpand\orcidicon}}
}
\usepackage{amsmath,amsfonts}
\usepackage{algorithmic}
\usepackage{algorithm}
\usepackage{array}

\usepackage{textcomp}
\usepackage{stfloats}
\usepackage{url}
\usepackage{verbatim}
\usepackage{graphicx}
\usepackage[numbers]{natbib} 

\usepackage[caption=false]{subfig}
\usepackage{subfloat}
\usepackage{caption}

\usepackage{amsthm}  

\theoremstyle{remark}  
\newtheorem{remark}{Remark}  

\begin{document}

\title{ISAC with Affine Frequency Division Multiplexing: An FMCW-Based Signal Processing Perspective}

\author{
Jiajun Zhu\hspace{-1.5mm}\orcidA{}, Yanqun Tang\hspace{-1.5mm}\orcidB{}, Cong Yi\hspace{-1.5mm}\orcidC{}, Haoran Yin\hspace{-1.5mm}\orcidD{}, \textit{Member, IEEE}, Yuanhan Ni\hspace{-1.5mm}\orcidE{}, \textit{Member, IEEE}, \\Fan Liu\hspace{-1.5mm}\orcidF{}, \textit{Senior Member, IEEE}, Zhiqiang Wei\hspace{-1.5mm}\orcidG{}, \textit{Member, IEEE}, and H{\"u}seyin Arslan\hspace{-1.5mm}\orcidH{}, \textit{Fellow, IEEE}


\thanks{
This work was supported in part by the Shenzhen Science and Technology Major Project under Grant KJZD20240903102000001 and in part by the Science and Technology Planning Project of Key Laboratory of Advanced IntelliSense Technology, Guangdong Science and Technology Department under Grant 2023B1212060024.
The work of Fan Liu was supported in part by the National Natural Science Foundation of China (NSFC) under Grant 62522107 and Grant 62331023.
This work of Zhiqiang Wei was supported in part by the Qin Chuang Yuan High-Level Innovation and Entrepreneurship Talent Program under Grant QCYRCXM-2023-094. (Corresponding author: Yanqun Tang.)

Jiajun Zhu, Yanqun Tang, Cong Yi and Haoran Yin are with the School of Electronics and Communication Engineering, Sun Yat-sen University, Shenzhen 518107, China, and also with the Guangdong Provincial Key Laboratory of Sea-Air-Space Communication, Shenzhen 518107, China (e-mail: zhujj59@alumni.sysu.edu.cn, tangyq8@mail.sysu.edu.cn, yicong@mail2.sysu.edu.cn, yinhr6@mail2.sysu.edu.cn).


Yuanhan Ni is with the School of Electronic and Information Engineering, Beihang University, Beijing 100191, China (e-mail: yuanhanni@buaa.edu.cn).

Fan Liu is with the National Mobile Communications Research Laboratory, Southeast University, Nanjing 210096, China (e-mail:
fan.liu@seu.edu.cn).

Zhiqiang Wei is with the School of Mathematics and Statistics, Xi’an
Jiaotong University, Xi’an, 710049, China (e-mail: zhiqiang.wei@xjtu.edu.cn).

H{\"u}seyin Arslan is with the Department of Electrical and Electronics Engineering, Istanbul Medipol University, Istanbul 34810, Turkey (email: huseyinarslan@medipol.edu.tr).
}}

\maketitle

\begin{abstract}

This paper investigates the sensing potential of affine frequency division multiplexing (AFDM) in high-mobility integrated sensing and communication (ISAC) from the perspective of radar waveforms.
We introduce an innovative parameter selection criterion that establishes a precise mathematical equivalence between AFDM subcarriers and Nyquist-sampled frequency-modulated continuous-wave (FMCW). This connection not only provides a clear physical insight into AFDM's sensing mechanism but also enables a direct mapping from the DAFT index to delay-Doppler (DD) parameters of wireless channels.
Building on this, we develop a novel input-output model in a DD-parameterized DAFT (DD-DAFT) domain for AFDM, which explicitly reveals the inherent DD coupling effect arising from the chirp-channel interaction. Subsequently, we design two matched-filtering sensing algorithms. The first is performed in the time-frequency domain
with low complexity, while the second is operated in the DD-DAFT domain to precisely resolve the DD coupling. 
Simulations show that our algorithms achieve effective pilot-free sensing and demonstrate a fundamental trade-off between sensing performance, communication overhead, and computational complexity. 
The proposed AFDM outperforms classical AFDM and other variants in most scenarios.
\end{abstract}

\section{Introduction}
\label{sec:introduction}
\IEEEPARstart{I}{ntegrated} sensing and communication (ISAC) has emerged as an innovative paradigm that combines sensing and communication functions within a single hardware platform and common signal framework \cite{liu2022integrated}. This convergence enables spatiotemporal resource reuse, unlocking the joint optimization of computation and storage resources. Early time-division ISAC prototypes are already being tested in 5G advanced networks, and the international telecommunication union radiocommunication sector has formally recognized ISAC as a cornerstone technology for IMT-2030 systems, highlighting its strategic importance for next-generation wireless ecosystems \cite{ITUM2516}. 
Nevertheless, time-division approaches inevitably waste temporal resources and introduce additional latency. As a result, a central research focus is enabling simultaneous communication and sensing within the same time-frequency (TF) resources on a unified system.

Waveform design is the key to ISAC system development, as it fundamentally delimits the performance envelope and the synergistic potential of the dual functions \cite{zhangOverviewSignalProcessing2021}. 
An ideal ISAC waveform should possess intrinsic flexibility, allowing for dynamic trade-offs between communication and sensing performance across diverse operational scenarios. 
For static environments, orthogonal frequency division multiplexing (OFDM) based ISAC waveform has been demonstrated to have significant advantages for monostatic ranging \cite{du2024reshaping,liu2025cpofdm}, which follows a communication-centric ISAC waveform design philosophy \cite{ayoub2025toward}.
Currently, high-speed mobility scenarios such as vehicular, unmanned aerial vehicle (UAV), and high-speed rail systems pose new challenges to communication waveforms. High mobility introduces large Doppler shifts and high channel dynamics, where significant Doppler effects cause severe inter-carrier interference in OFDM systems, degrading their communication reliability \cite{bemani2023affineb}. Meanwhile, maintaining accurate channel estimation in such dynamic channels requires denser pilot symbols, which reduces communication efficiency \cite{li2000pilotsymbolaided}. For sensing, velocity estimation typically requires exploiting multiple OFDM symbols, imposing stricter formatting requirements and reducing adaptability to scenarios with rapidly-moving targets \cite{sturm2011waveform,zhang2024OFDM}. Considering the need for resource efficiency and higher sensing rate in next-generation wireless networks, achieving dual functions within a single-symbol framework has become a promising research focus.

Specifically, crafting a communication-centric ISAC waveform and its associated processing framework that delivers excellent sensing ranging and velocity estimation performance while remaining communication resource efficient is of great importance \cite{zhangEnablingJointCommunication2022}.
To overcome OFDM's shortcoming in Doppler sensitivity, alternative waveforms like orthogonal time frequency space (OTFS) \cite{hadani2017orthogonal} and affine frequency division multiplexing (AFDM) \cite{bemani2023affineb,yin2025afdmwc,zhouAFDM} have been proposed. OTFS is a pulse-train based communication waveform, which leverages the properties of pulse radar to achieve Doppler robustness for communications and accurate range/velocity estimation for sensing. Moreover, AFDM presents a compelling alternative by employing orthogonal chirp-based subcarriers generated via the discrete affine Fourier transform (DAFT). 
Its tunable chirp parameters unlock full path separation capabilities under high mobility, surpassing and encompassing orthogonal chirp division multiplexing (OCDM) as a specific case of this broader AFDM framework\cite{bemani2023affineb}.
Furthermore, studies reveal that AFDM can outperform OTFS in diversity gain, channel estimation efficiency, and compatibility with OFDM \cite{bemani2023affineb,yinDiagonallyReconstructedChannel2024}, suggesting its coexistence and suitability for future wireless networks without troublesome transition from OFDM \cite{rouAFDM6G}. 
The source of these advantages lies in AFDM subcarriers whose inherent chirp characteristics are intrinsically aligned with those of radar waveforms. This alignment, coupled with flexible parameterization, renders AFDM a highly promising ISAC candidate. Notably, inspired by the advantages of chirp signals, recent works have also begun integrating chirp features with OTFS to boost its ISAC performance \cite{bondre2022dualuse,zegrar2024novel,ubadah2025zakotfs}.

Recent studies have highlighted the growing potential of AFDM in ISAC, with research branching into various facets of system design and signal processing \cite{yin2025afdmisacmag,liu2025afdm,liu2025afdmim}. For instance, some works have established foundational frameworks for performance analysis, examining the impact of AFDM parameters using multi-symbol frames \cite{niIntegratedSensingCommunications2025b} or focusing on the ambiguity function (AF) of AFDM signals \cite{yin2025ambiguity,ni2025ambiguity}. The design of sensing methodologies has also been a key area of investigation. One prominent approach involves using deterministic pilots, with studies proposing specific pilot-assisted waveforms and formulating the corresponding AF \cite{zhang2025afdmenabled}. In contrast, another line of research has pursued blind sensing, achieving parameter estimation without prior symbol knowledge in bistatic radar scenarios \cite{ranasinghe2024blind}. Furthermore, research has explored specialized system configurations and applications \cite{ZSui}, such as implementing full-duplex ISAC with efficient self-interference suppression \cite{lu2025daftdomain} and developing preamble-based frame structures for sensing-aided channel estimation in bistatic environments \cite{zhuAFDMBasedBistaticIntegrated2024}.

However, these studies primarily analyze the direct relationship between AFDM parameters and sensing performance metrics, such as the AF or the Cramér-Rao bound (CRB) \cite{baoPerformanceTradeoffCommunication2024, bedeer2025ambiguitya}.
This is attributed to the fact that these analysis being a mathematical abstraction, lacks examination of the physical meaning inherent in AFDM, limiting the generality of its conclusions. For example, the selected parameters may no longer be applicable when other performance metrics are adopted. In contrast, a waveform-centric perspective enables the analysis of intrinsic physical meaning independent of specific performance metrics. Clarifying the physical meaning not only ensures that waveform designs are physically realizable, but also facilitates their extension to other applications. 
More importantly, current AFDM research focuses solely on the characteristics of chirp, which fails to fully leverage AFDM's advantages since its inherent multi-chirp nature remains untapped. Furthermore, as one of two important waveforms capable of full diversity performance under time-varying channels, the fair comparison between AFDM and OTFS has drawn significant attention. Therefore, there is an urgent need to unify them under a common framework, which relies on a deeper interpretation of the physical meaning of AFDM. For example, identifying a widely recognized signal as the most fundamental unit of AFDM, whose structure and TF characteristics have been thoroughly studied, enables all properties of AFDM to be derived from this basic signal.


Inspired by the insight that OTFS subcarriers are equivalent to radar pulse trains \cite{bondre2022dualuse,zhang2023radar}, we adopt a signal processing perspective rooted in the subcarriers of AFDM. 
The core motivation of this paper is to leverage the parameter flexibility of AFDM modulation to establish a connection between AFDM and frequency-modulated continuous-wave (FMCW) and then design effective ISAC receiver. 
Compared to conventional AFDM, the newly configured AFDM subcarriers can leverage the periodic characteristics of chirps, i.e., those of FMCW signals, significantly enhancing the matching gain in pulse-accumulation-based algorithms. And it is precisely these periodic chirps that form the most fundamental units of the proposed AFDM, providing its core physical meaning.
We further analyze the physical meanings in AFDM, such as its TF properties similar to the FMCW waveform and the delay-Doppler (DD) properties which enable efficient radar sensing.
Furthermore, we focus on developing effective algorithms based on a single AFDM symbol, enabling a dynamic trade-off between communication throughput and sensing performance without requiring conventional complicated multi-symbol processing \cite{xiongFundamentalTradeoffIntegrated2023,sumer2025novel}. 
The main contributions of this paper are summarized as follows:
\begin{itemize}
    \item We reinterpret AFDM from an FMCW radar perspective, a viewpoint hindered by conventional parameter selections. To achieve this, we propose a novel parameter selection criterion that forges a precise mathematical equivalence between AFDM subcarriers and Nyquist-sampled periodic chirp waveforms. This framework not only provides a clear physical intuition for the AFDM sensing mechanism but also establishes a fundamental comparison to OTFS, distinguishing AFDM's continuous periodic chirps from the periodic delta pulses of pulse-Doppler radar.

    \item We introduce a novel DD parameterization mapped from the DAFT index to analyze AFDM waveform from a more intrinsic, channel-centric viewpoint. This allows us to derive a new input-output relationship in what we term the DD-parameterized DAFT (DD-DAFT) domain via discrete periodic AF. This relationship is pivotal as it explicitly reveals the inherent DD coupling effect of chirp-based sensing, and models the channel interaction as a two-dimensional convolution. This formulation clarifies the structural similarities and key differences between the DD-DAFT domain of AFDM and the DD domain of OTFS.

    \item Leveraging the FMCW characteristics revealed in the above two domains, we first propose a TF-domain monostatic sensing algorithm using fast Fourier transform-based matched filtering, simplifying to dechirp when pilot considered. To overcome inherent delay-Doppler coupling, we develop an DD-DAFT domain algorithm that explicitly decouples these effects. Both methods are applicable to pilot-free AFDM.

\end{itemize}

The remainder of this paper is organized as follows. 
Section \ref{sec:preliminaries} introduces the system model.
In Section \ref{sec:relationship}, we establish the relationship between any AFDM subcarrier and FMCW signal.
Based on this, we introduce a new AFDM input-output relationship with DD parameters in Section \ref{sec:ddafdm}. 
And in Section \ref{sec:algorithms}, we propose the sensing algorithms.
Simulation results are given in Section \ref{sec:simulation}. Section \ref{sec:conclusion} concludes the paper.

\section{Preliminaries}
\label{sec:preliminaries}
\subsection{ISAC Channel Model}
This paper employs a channel model rooted in the physical geometry of scatterers \cite{Zhoumodeling}. The model utilizes the physical quantities of delay and Doppler shift to characterize each resolvable propagation path. Each path, originating from a scatterer, can also be interpreted as a sensing target. Specifically, the delay $\tau$ corresponds to the signal propagation time, which is directly related to the distance of the scatterer. The Doppler shift $\nu$ represents the signal frequency deviation caused by relative motion between the transceiver and the scatterer, which is directly related to the radial velocity of the scatterer. This modeling approach, based on geometric parameters, accurately captures the time-varying characteristics of wireless channels, proving particularly effective for systems with high frequencies, wide bandwidths, and long observation times. A key parallel exists with radar systems, where the core task of measuring target range and velocity aligns directly with estimating the delay and Doppler parameters of channel. Consequently, radar sensing can be realized by extracting these channel parameters from the received signal.
\begin{figure}[t]
    \centering
    \includegraphics[width=0.8\linewidth]{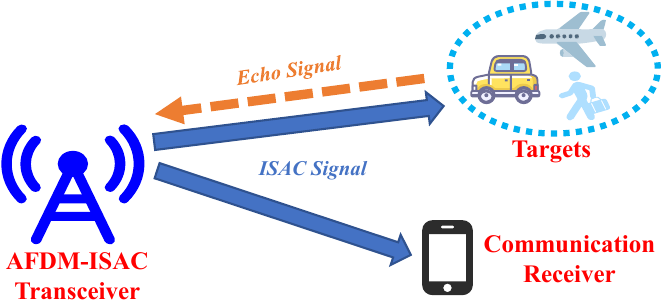}
    \caption{The AFDM-based ISAC scenario.}
    \label{fig:AFDM_ISAC_channel}
    \vspace{-10pt}
\end{figure}
In a representative ISAC scenario, an ISAC transceiver transmits a unified signal, as depicted in Fig. \ref{fig:AFDM_ISAC_channel}. This signal is received by a user terminal for communication and simultaneously reflected by targets back to the transceiver to enable monostatic active sensing.

Therefore, a single, unified time-varying channel impulse response model based on discrete scatterers can effectively describe signal propagation in ISAC systems:
\begin{equation}
  g(t,\tau)=\sum_{i=1}^P h_i e^{-j2\pi \nu_it}\delta (\tau-\tau_i), \label{TDchannel}
\end{equation}
where $P$ is the total number of resolvable scatterers (paths) in the environment. $h_i$, $\tau_i$, and $\nu_i$ represent the complex path gain (including path loss, scattering/reflection coefficients, and phase changes), delay, and Doppler shift of the $i$-th scatterer, respectively. The $e^{-j2\pi \nu_it}$ term represents time-varying phase rotation due to Doppler shift, and $\delta (\tau-\tau_i)$ describes the delay of the $i$-th path.

In monostatic ISAC scenarios, the signal experiences a round trip, and the corresponding delay $\tau_i$ and Doppler shift $\nu_i$ are given by $
	\tau _i=\frac{2R_{i}}{V_\mathrm{c}},\quad 	\nu_i=\frac{2v_{i}f_c}{V_\mathrm{c}}
$, where $V_\mathrm{c}$ is the speed of light, $R_{i}$ is the one-way range between the transceiver and the target, $v_{i}$ is the radial velocity of the target, and $f_c$ is the carrier frequency of the signal.
A key property of this geometric channel model is that in typical wireless environments, the position and velocity of scatterers change slowly relative to symbol or frame durations \cite{shi2021deterministic}. This allows the channel parameters $h_i$, $\tau_i$, and $\nu_i$ to be considered quasi-static throughout an ISAC signal frame, which is often on the order of milliseconds \cite{paier2008nonwssus}.

For digital signal processing, the continuous-time channel model in \eqref{TDchannel} requires discretization. Given a system signal bandwidth $B$ and a total observation time $T$, the Nyquist sampling theorem and the duality between time and frequency domains dictate that the channel delay resolution is approximately $1/B$ and the Doppler resolution is approximately $1/T$. Accordingly, the continuous delay $\tau_i$ and Doppler shift $\nu_i$ can be mapped to discrete delay and Doppler taps, denoted by $l_i$ and $k_i$, respectively:
\begin{equation}
   l_i=\mathrm{round}( B \tau_i ),\quad {k_i}=\mathrm{round}({T\nu_i)},\label{rangesample}
\end{equation}
where $\mathrm{round}( \cdot )$ is the round operation. The delay tap $l_i$ typically lies in the range $0 \le l_i \le l_{\mathrm{max}}$, where $l_{\mathrm{max}} = \mathrm{round}( B\tau_{\mathrm{max}})$ and $\tau_{\mathrm{max}}=R_{\mathrm{max}}/V_\mathrm{c}$ is the maximum delay of interest, corresponding to a maximum distance $R_{\mathrm{max}}$. Similarly, the Doppler tap $k_i$ falls within the range $k_i \in[-k_{\mathrm{max}},+k_{\mathrm{max}}]$, where $k_{\mathrm{max}}=\mathrm{round}( T\nu_{\mathrm{max}})$. Here, $\nu_{\mathrm{max}}=f_{\mathrm c}v_{\mathrm{max}}/V_\mathrm c$ is the maximum Doppler shift, corresponding to a maximum relative radial velocity $v_{\mathrm{max}}$.

Using these discrete taps, the effect of the $i$-th scatterer on the transmitted signal can be represented in the discrete time-delay domain. Assuming a discrete signal length of $N_{\mathrm{c}}=TB$. The total discrete time-varying channel impulse response is:
\begin{equation}
  g[n,l] = \sum_{i=1}^Ph_i e^{-j2\pi \frac{k_i n}{N_{\mathrm{c}}}}\delta[l-l_i], \label{eq:channel_model}
\end{equation}
where $n$ is the discrete time index, $l$ is the discrete delay tap index, and $P$ is the number of paths. It is important to note that different frame structures, involving changes in symbol duration or subcarrier spacing, will alter the discretization grid for delay and Doppler \cite{bello1963characterization}. Furthermore, practical systems often add a cyclic prefix (CP) of length $L_{\mathrm{CP}}$, making the actual symbol duration $(N_{\mathrm{c}}+L_{\mathrm{CP}})T_s$, where $T_s=1/B$. This extended duration must be accounted for in a precise analysis.
\subsection{AFDM Signal Model}
\label{cha:introduction}
AFDM is a multicarrier modulation scheme based on the DAFT that can be viewed as a generalization of OFDM.
The AFDM modulation process generates a time-domain signal $s[n]$ from information symbols $x[m]$ as follows:
\begin{equation}
s[n] = \frac{1}{\sqrt{N_{\mathrm{c}}}} \sum_{m=0}^{N_{\mathrm{c}}-1} x[m] e^{j 2\pi \left( c_1 n^2 + \frac{1}{N_{\mathrm{c}}} m n + c_2 m^2 \right)}.\label{eq:AFDM_modulation}
\end{equation}
where $n = 0, \dots, N_{\mathrm{c}}-1$ and $N_{\mathrm{c}}$ is the number of subcarriers. The primary distinction from OFDM is the inclusion of two additional chirp parameters, $c_1$ and $c_2$. These parameters define the chirp characteristics of the subcarriers Specifically, the term $e^{j 2\pi c_1 n^2}$ functions as a time-domain multiplication (chirp window), while $e^{j 2\pi c_2 m^2}$ acts as a frequency-domain multiplication or time-domain convolution (chirp filter). When $c_1$ and $c_2$ are both zero, AFDM simplifies to OFDM.
The basis function for the $m$-th subcarrier is given by:
\begin{equation}
\psi_m[n] = e^{j 2\pi \left( c_1 n^2 + \frac{1}{N_{\mathrm{c}}} m n + c_2 m^2 \right)}.\label{eq:afdm_subcarrier}
\end{equation}

This function is a complex exponential signal containing an additional quadratic phase term $c_1 n^2$. This term creates an instantaneous frequency that varies linearly with time $n$. Consequently, each basis function is a discrete chirp signal overlaid on a fundamental frequency of $m/N_{\mathrm{c}}$. A key property is that for any choice of $c_1$ and $c_2$, these subcarrier basis functions remain strictly orthogonal:
\begin{equation}
\begin{aligned}
    \langle \boldsymbol{\psi}_m, \boldsymbol{\psi}_{m'} \rangle = \sum_{n=0}^{N_{\mathrm{c}}-1} {\psi}_m[n] ({\psi}_{m'}[n])^* = N_{\mathrm{c}} \delta[m - m'].
\end{aligned} \label{eq:orthogonal}
\end{equation}

\begin{figure}[t]
    \centering
    \includegraphics[width=\linewidth]{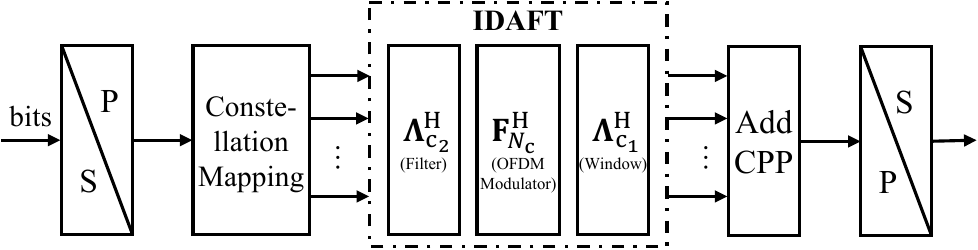}
    \caption{AFDM modulation block diagram.}
    \label{fig:AFDM_modulation_fig}
    \vspace{-10pt}
\end{figure}

As depicted in Fig. \ref{fig:AFDM_modulation_fig}, the IDAFT operation of AFDM modulation can be implemented in three steps. First, the input symbols $x[m]$ in the DAFT domain are pre-weighted, i.e, filtered, by a frequency domain chirp phase: $ \tilde{x}[m] = x[m] e^{j 2\pi c_2 m^2} $. Second, an $N_{\mathrm{c}}$-point inverse discrete Fourier transform (IDFT), i.e., OFDM modulation, is performed on the weighted symbols: $ \tilde{s}[n] = \text{IDFT}\{\tilde{x}[m]\}_n $. Finally, the resulting time-domain signal is post-weighted, i.e., windowed, by a time-domain chirp phase: $ s[n] = \tilde{s}[n] e^{j 2\pi c_1 n^2} $.

The core advantage of AFDM stems from the parameters $c_1$ and $c_2$, whose additional degrees of freedom enable flexible adaptation to different channel environments \cite{yin2025afdmwc}.
Similar to OFDM, AFDM requires a guard interval to eliminate inter-symbol interference. However, due to the time-domain chirp factor $ e^{j 2\pi c_1 n^2} $, the AFDM signal is not strictly periodic but rather quasi-periodic, satisfying $ s[n + k N_{c}] = e^{j 2\pi c_1 (k^2 N_{c}^2 + 2 k N_{c} n)} s[n] $. Consequently, AFDM employs a chirp periodic prefix (CPP) of length $L_\mathrm{CPP}>l_\mathrm{max}$ as the guard interval. It not only copies the signal tail to the head but also multiplies by a specific compensating phase factor to ensure the periodicity of the signal with CPP added in the DAFT, given by:
\begin{equation}
s[n] = e^{-j 2\pi c_1 (N_{c}^2 + 2 N_{c} n)} s[n + N_{c}], \quad n = -L_{\text{CPP}}, \dots, -1. \label{eq:cppafdmmodulate}
\end{equation}

At the ISAC receiver, after removing the CPP, we obtain the corresponding echo signal as:
\begin{equation}
    r[n] = \sum_{i=1}^P h_i s[(n - l_i)_{N_{\mathrm{c}}}] e^{-j 2 \pi \frac{k_i n}{N_{\mathrm{c}}}} + w[n].\label{eq:afdm_echo}
\end{equation}
where $w[n]$ is the addictive white Gaussian noise (AWGN), $(\cdot)_n$ denotes the modulo $n$ operation, this is because the transmitted signal with CPP, after removing the CPP at the receiver, results in a remaining portion that is equivalent to a cyclic shift version of the original signal.

\subsection{FMCW Signal Model}
FMCW radar transmits a continuous waveform, in contrast to traditional pulsed radar systems. The transmitted signal is typically a periodic sequence of linear frequency modulation (LFM) waveforms, commonly referred to as chirps. Owing to their large time–bandwidth product, chirp signals enable high range resolution through pulse compression at the receiver while maintaining low peak transmit power.
A representative baseband up-chirp signal can be expressed as:
\begin{equation}
s_{\text{chirp}}(t) = e^{j 2\pi \left( f_0 t + \frac{\mu}{2} t^2 \right)}, \quad 0 \le t < T,
\end{equation}
where $f_0$ is the start frequency, $f_1$ is the end frequency, $T$ is the chirp duration or sweep period, and $\mu = (f_1 - f_0)/T$ is the frequency modulation rate. The instantaneous frequency, $f(t) = f_0 + \mu t$, increases linearly with time. The occupied signal bandwidth is $B = |f_1 - f_0| = |\mu| T$.

In FMCW radar applications, these chirp signals are transmitted periodically. If the signal is transmitted $K$ times with a period $T$, it forms a periodic chirp waveform with a total duration of $KT$, expressed as:
\begin{equation}
s_{\text{FMCW}}(t) = \sum_{q=0}^{K-1} s_{\text{chirp}}(t - qT). 
\end{equation}

For digital implementation, the signal is sampled at the Nyquist rate $f_s = B$, corresponding to a sampling interval of $T_s = 1/B$. In this paper, we assume a start frequency $f_0 = 0$, which implies $f_1 = B$ and $\mu = B/T$. Let $N_{\mathrm{p}} = T/T_s = TB$ be the number of samples within a single chirp period. The discrete time representation of a single chirp signal is given by:
\begin{equation}
s_{\text{chirp}}[n] = e^{j \pi \frac{n^2}{N_{\mathrm{p}}}},0 \le n \le N_{\mathrm{p}}-1.
\end{equation}

Then the complete discrete FMCW signal, for $0 \le n < N_{\mathrm{c}},N_{\mathrm{c}}=KN_{\mathrm{p}}$, comprising $K$ periodic chirps, is given by:
\begin{equation}
s_{\text{FMCW}}[n] = \sum_{q=0}^{K-1} e^{j \pi \frac{(n - qN_{\mathrm{p}})^2}{N_{\mathrm{p}}}} \operatorname{Rect}\left[ \frac{n - qN_{\mathrm{p}}}{N_{\mathrm{p}}} \right],\label{eq:FMCW_discrete_signal}
\end{equation}
where $ \operatorname{Rect}[x/N_{\mathrm{p}}] $ denotes a rectangular window of length $ N_{\mathrm{p}} $.
Ignoring noise for simplicity, the corresponding echo signal can be modeled as:
\begin{equation}
    r_{\text{FMCW}}[n] = \sum_{i=1}^P h_i s_{\text{FMCW}}[n - l_i] e^{-j 2 \pi \frac{k_i n}{N_{\mathrm{c}}}}.\label{eq:fmcw_echo}
\end{equation}

Compared to \eqref{eq:afdm_echo}, this equation lacks the modulo operation. To give the FMCW echo a cyclic shift characteristic, we add a CPP to the FMCW signal in the subsequent section.

\section{The Relationship Between AFDM and Chirp Waveforms}
\label{sec:relationship}

\subsection{Parameters Selection}
\label{subsec:parameter_selection}
AFDM possesses sufficient degrees of freedom in parameter selection to potentially achieve excellent characteristics like periodic chirps. Here, the derivation begins by considering the $0$-th subcarrier, $\psi_0[n]$. For periodicity with period $N_{\mathrm{p}} \in \mathbb{Z}^+$, the signal must satisfy the condition that for any integer $q$, $\psi_0[n + qN_{\mathrm{p}}]$ differs from $\psi_0[n]$ only by a constant phase factor. According to the definition in \eqref{eq:afdm_subcarrier}, we have:
\begin{equation}
\begin{aligned}
\psi_0[n + qN_{\mathrm{p}}] &= e^{j 2 \pi c_1 (n + qN_{\mathrm{p}})^2} = \psi_0[n] e^{j 2 \pi c_1 (2 q N_{\mathrm{p}} n + q^2 N_{\mathrm{p}}^2)}.
\end{aligned}
\end{equation}

For strict periodicity, the phase term $c_1 (2 q N_{\mathrm{p}} n + q^2 N_{\mathrm{p}}^2)$ must be an integer for all integers $n$ and $q$. This requirement imposes two conditions. First, the term linear in $n$, $2 c_1 q N_{\mathrm{p}} n$, must be an integer, which implies that $2 c_1 N_{\mathrm{p}}$ must be an integer. We denote this integer by $Z_A \in \mathbb{Z}^+$, leading to:
\begin{equation}
c_1 = \frac{Z_A}{2 N_{\mathrm{p}}}.
\end{equation}

Second, the term quadratic in $q$, $c_1 q^2 N_{\mathrm{p}}^2$, must also be an integer. Substituting the expression for $c_1$ yields $\frac{Z_A N_{\mathrm{p}}}{2} q^2$. For this formula to hold for all $q$, the product $Z_A N_{\mathrm{p}}$ must be an even number, which can be expressed as:
\begin{equation}
Z_A N_{\mathrm{p}} \equiv 2\mathbb{Z}^+.
\end{equation}

This parity constraint dictates the choice of $Z_A$:
\begin{equation}
Z_A \in
\begin{cases}
2\mathbb{Z}^+, & \text{if } N_{\mathrm{p}} \text{ is odd}, \\
\mathbb{Z}^+, & \text{if } N_{\mathrm{p}} \text{ is even}. 
\end{cases}\label{eq:ZA_Np_selection}
\end{equation}

Assuming the AFDM symbol length $N_{\mathrm{c}}$ is an integer multiple of the chirp period $N_{\mathrm{p}}$, such that $N_{\mathrm{c}} = K N_{\mathrm{p}}$ for some integer $K \ge 1$, we can express $c_1$ in terms of $N_{\mathrm{c}}$:
\begin{equation}
c_1 = \frac{Z_A K}{2 N_{\mathrm{c}}}. 
\end{equation}

To specify a unique value for $Z_A$, we analyze the instantaneous frequency of the base chirp $\tilde{\psi}_0[n] = e^{j 2 \pi \frac{Z_A}{2 N_{\mathrm{p}}}n^2}$ for $0\le n < N_{\mathrm{p}}$. The frequency, $f(n) \approx \frac{Z_A}{N_{\mathrm{p}}} n$, sweeps from $0$ to $Z_A ( 1 - 1/N_{\mathrm{p}} )$. To ensure the fundamental period is exactly $N_{\mathrm{p}}$, rather than a sub-multiple, we set $Z_A = 1$. This choice requires $N_{\mathrm{p}}$ to be even and fixes the value of $c_1$:
\begin{equation}
c_1 = \frac{1}{2 N_{\mathrm{p}}} = \frac{K}{2 N_{\mathrm{c}}}. \label{eq:c1_selection}
\end{equation}

While other integer values for $Z_A$ satisfying the parity constraint are possible, corresponding to faster chirp sweeps that cover the bandwidth multiple times within the period $N_{\mathrm{p}}$, the choice of $Z_A=1$ represents the most fundamental case, creating a base chirp that sweeps the bandwidth exactly once. In addition, a small $Z_A$ minimizes the overhead of channel estimation as much as possible. Based on this canonical choice, $Z_A$ can be readily extended to other values. 
Considering that $N_{\mathrm{c}}$ is typically a power of two in practical multicarrier systems and to satisfy the path separability requirement \cite{bemani2023affineb} of $K \ge 2k_{\max} + 1$, the value of $K$ is defined as:
\begin{equation}
    K = 2^{\lceil \log_2(2k_{\max} + 1) \rceil},
\end{equation}
where $\lceil\cdot \rceil$ denotes the ceiling operation.
Furthermore, a consequence of $N_{\mathrm{p}}$ being even is that $N_{\mathrm{c}} = K N_{\mathrm{p}}$ must also be even. This allows the CPP defined in \eqref{eq:cppafdmmodulate} to be simplified to a standard CP. The selection of the parameter $c_2$ is justified in the first two paragraphs of \eqref{eq:subcarrier_and_chirp} in Subsection \ref{subsec:chirp_AFDM} , where it is shown that setting $c_2 = 0$ is necessary to eliminate the $m^2$-introduced phase difference between the FMCW echo and the AFDM subcarrier, so here’s the conclusion up front:
\begin{equation}
    c_2 = 0. \label{eq:c2_selection}
\end{equation}

With the parameters chosen according to \eqref{eq:c1_selection} and \eqref{eq:c2_selection}, the $0$-th AFDM subcarrier becomes a perfectly periodic chirp signal. This differs from classic AFDM with specified parameters designed for communication and can be viewed as a special case designed for ISAC (e.g., to limit communication performance degradation \cite{safdm} while optimizing sensing performance) within the AFDM framework \cite{bemani2023affineb}). This transformation is pivotal, as it allows for the direct application of signal processing techniques from traditional FMCW radar to AFDM-based sensing.

Fig. \ref{fig:ali_afdm_waveform} and Fig. \ref{fig:pro_afdm_waveform} illustrate the effect of this parameter selection. It contrasts the time-domain waveforms and TF distributions of AFDM signals generated with classic parameters \cite{bemani2023affineb} against those generated with the proposed parameters\footnote{Notably, the proposed parameter selection adheres to the design criteria in \cite{bemani2023affineb}, preserving AFDM's orthogonality and path resolution. It satisfies the three requirements established therein: i) orthogonality (independent of $c_1, c_2$ per \eqref{eq:orthogonal}), ii) the $c_1 > \frac{2k_{\max}}{l_j - l_i}$ condition for path distinction, and iii) the rational $c_2 \ll \frac{1}{2N_{\mathrm{c}}}$ constraint. To ensure ii), a relatively larger $c_1$ is adopted, which involves $2(K-2k_{\max}-1)(l_{\max}+1)$ additional guard interval samples.}. The proposed configuration clearly produces a periodic chirp structure for the $0$-th subcarrier, which is selected as the foundational basis for our analysis. The rationale for this choice is that under these parameters, each constituent chirp segment $\tilde \psi_0[n],0\le n \le N_{\mathrm{p}}-1$ of $\psi_0[n],0\le n \le N_{\mathrm{c}}-1$ spans the entire normalized digital frequency range of $[0, 1)$, a behavior analogous to conventional FMCW radar signals. Other subcarriers ($m \neq 0$) possess non-zero starting frequencies $m/N_{\mathrm{c}}$ at the starting time sample $n=0$ and can not cover the bandwidth continuously. 
The $0$-th subcarrier thus provides the clearest analogy to a traditional FMCW radar signal, and as shown, all other subcarriers can be constructed from it.

\clearpage
\subsection{FMCW-Based AFDM Subcarrier Construction}
\label{subsec:chirp_AFDM}


\begin{figure*}[htbp]
	\centering
	\subfloat[AFDM under classic parameters ($0$-th subcarrier)]{\includegraphics[width=0.48\linewidth]{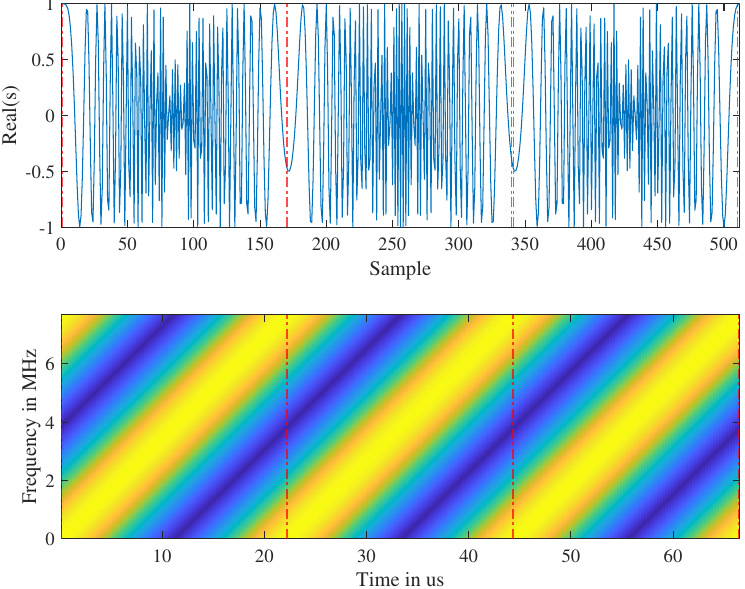}\label{fig:ali_afdm_waveform}}
	\subfloat[AFDM under proposed parameters ($0$-th subcarrier)]{\includegraphics[width=0.48\linewidth]{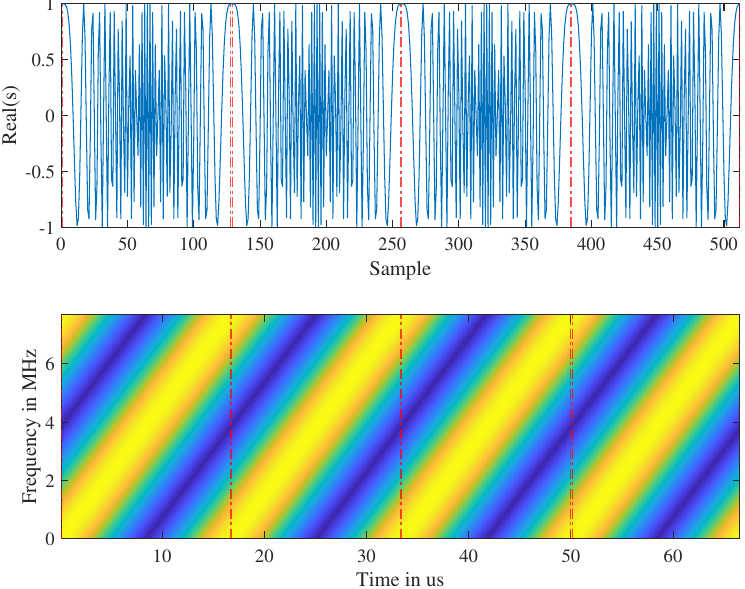}\label{fig:pro_afdm_waveform}}
    
	\subfloat[Proposed AFDM's echo with delay tap=10 (($N_{\mathrm{c}}$-$10K$)-th subcarrier)]{\includegraphics[width=0.48\linewidth]{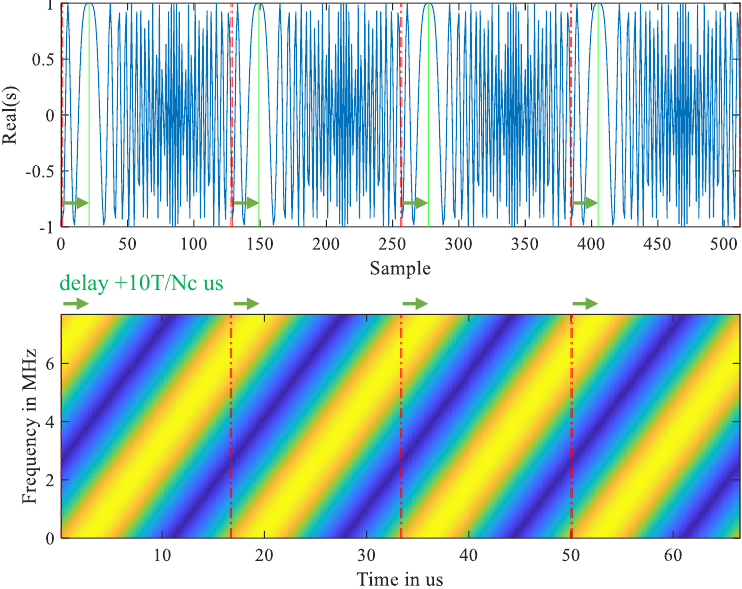}\label{fig:pro_afdm_waveform_de}}
	\subfloat[Proposed AFDM's echo with Doppler tap=1 (($N_{\mathrm{c}}$-$1$)-th subcarrier)]{\includegraphics[width=0.48\linewidth]{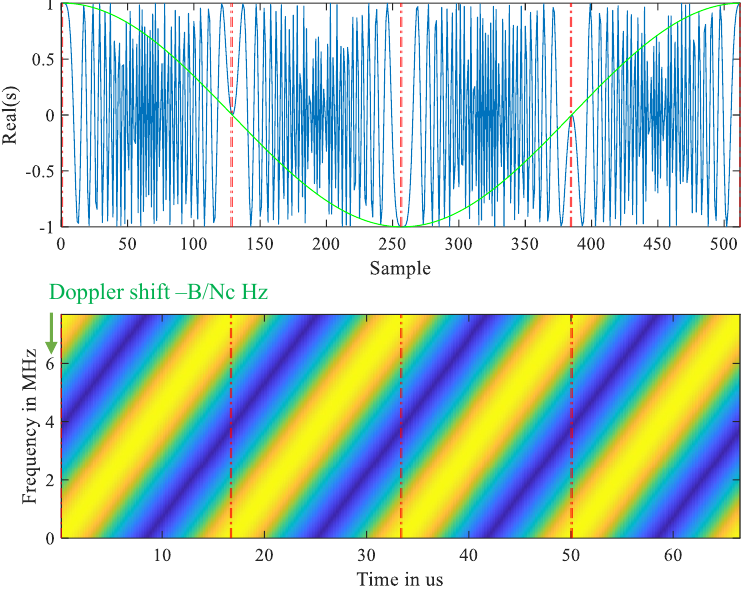}\label{fig:pro_afdm_waveform_do}}
	\caption{Time domain waveforms and TF distributions of the $0$-th AFDM subcarrier and its echos.}
	\label{AFDM_waveform}
    \vspace{-10pt}
\end{figure*}

This subsection formalizes the structure of the AFDM subcarriers using the parameters selected in Subsection \ref{subsec:parameter_selection}. With $c_1$ and $c_2$ defined by \eqref{eq:c1_selection} and \eqref{eq:c2_selection}, the $m=0$ subcarrier is given by:
\begin{equation}
\psi_0[n] = e^{j 2\pi c_1 n^2} = e^{j 2\pi \frac{n^2}{2N_{\mathrm{p}}}}, \quad n=0, \dots, N_{\mathrm{c}}-1.
\end{equation}

This expression represents a periodic extension of a base chirp, $\tilde{\psi}_0[n] = e^{j \pi n^2 / N_{\mathrm{p}}}$ for $0 \le n < N_{\mathrm{p}}$, with phase continuity ensured at the segment boundaries. Formally, $\psi_0[n]$ is equivalent to $K$ concatenated instances of this base chirp, matching the structure of the FMCW signal in \eqref{eq:FMCW_discrete_signal}:
\begin{equation}
\begin{aligned}
\psi_0[n] &= \sum_{q=0}^{K-1} \tilde{\psi}_0[(n - qN_{\mathrm{p}})_{N_{\mathrm{p}}}] \\
&= \sum_{q=0}^{K-1} e^{j \pi \frac{(n-qN_{\mathrm{p}})^2}{N_{\mathrm{p}}}} \operatorname{Rect}\left[\frac{n-qN_{\mathrm{p}}}{N_{\mathrm{p}}}\right].
\end{aligned} \label{eq:periodic_chirp_construction}
\end{equation}

This confirms that the $0$-th AFDM subcarrier is mathematically identical to a discrete periodic chirp (i.e., FMCW) signal:
\begin{equation}
    \begin{aligned}
        &\tilde \psi_0[n]=s_{\text{chirp}}[n],0 \le n \le N_{\mathrm{p}}-1\\
        &\psi_0[n]=s_{\text{FMCW}}[n],0 \le n \le N_{\mathrm{c}}-1. \label{eq:sub0_and_chirp}
    \end{aligned}
\end{equation}

Next, we show that any subcarrier $\psi_m[n]$ is equivalent to an echo of this base signal. An echo of $\psi_0[n]$ from a channel path with delay $l$ and Doppler index $k$ can be modeled as:
\begin{equation}
\begin{aligned}
r[n] &= \psi_0[(n - l)_{N_{\mathrm{c}}}] e^{-j 2\pi \frac{k}{N_{\mathrm{c}}} n} \\
&= e^{j \pi \frac{(n-l)^2}{N_{\mathrm{p}}}} e^{-j 2\pi \frac{k}{N_{\mathrm{c}}} n} \\
&= e^{j 2\pi (\frac{n^2}{2N_{\mathrm{p}}})} e^{-j 2\pi \frac{n}{N_{\mathrm{c}}} (Kl + k)} e^{j \pi \frac{l^2}{N_{\mathrm{p}}}}.
\end{aligned} \label{subcarrier_echo}
\end{equation}

Comparing the echo in \eqref{subcarrier_echo} with the general form of $\psi_m[n] = e^{j 2\pi (\frac{n^2}{2N_{\mathrm{p}}} + \frac{mn}{N_{\mathrm{c}}}+c_2m^2)}$, the quadratic phase terms are identical. Equating the linear phase terms yields $e^{-j2\pi\frac{m}{N_c}} = e^{-j2\pi\frac{Kl+k}{N_c}}$, i.e., $Kl+k+m \equiv (0)_{N_c}$. Given the subcarrier index $m\in[0,N_{\mathrm{c}}-1]$ and the DD coordinate $(l,k)$, where $l \in [0, N_{\mathrm{p}}-1]$ and $k \in [0, K-1]$, the sum is bounded as $Kl+k+m\in [0, 2N_{\mathrm{c}}-2]$. So $Kl+k+m$ can only yields two valid solutions, i.e., $0$ and $N_{\mathrm{c}}$, which can be described as:
\begin{equation}
\begin{aligned}
	m = (-Kl - k)_{N_{\mathrm{c}}} &= (N_{\mathrm{c}} - Kl - k)_{N_{\mathrm{c}}}\\
	&= \begin{cases} 
	0, &  l = k = 0 \\ 
	N_{\mathrm{c}} - Kl - k, & \text{otherwise}
	\end{cases},
\end{aligned} \label{eq:mapping}
\end{equation}
where the range of $k \in [0, K-1]$ is equivalent to  $k \in [-K/2, K/2-1]$ due to the periodicity of the exponential function $e^{-j 2\pi \frac{k}{N_{\mathrm{c}}} n}$. 

Finally, to maintain a consistent structure, the constant phase terms must be reconciled. The AFDM subcarrier contains the term $c_2m^2$, while the echo contains a phase term dependent on the delay, $e^{j\pi l^2/N_{\mathrm{p}}}$. A direct equivalence without introducing complex dependencies is achieved by setting $c_2=0$, which validates the choice made in \eqref{eq:c2_selection}.
This choice is crucial for enabling practical sensing. A non-zero $c_2$ would introduce a symbol-dependent phase term $e^{j2\pi c_2m^2}$ 
that becomes intricately coupled with the DD parameters of targets in channel. Such a dependency would complicate the MF design significantly, potentially requiring prior knowledge of the transmitted subcarriers or rendering the sensing process impractical. Therefore, setting $c_2$ is a necessary condition for achieving the clean,  subcarrier-independent sensing capability that underpins our proposed sensing algorithms.

With this complete parameter set, substituting the mapping from \eqref{eq:mapping} into the subcarrier definition $\psi_m[n] = e^{j 2\pi \left( c_1 n^2 + \frac{1}{N_{\mathrm{c}}} m n +c_2m^2\right)}$ yields a critical relationship with \eqref{subcarrier_echo}:
\begin{equation}
\begin{aligned}
	\psi_{m=(N_{\mathrm{c}} - Kl - k) _{N_{\mathrm{c}}}}[n] &= \psi_0[(n - l)_{N_{\mathrm{c}}}] e^{-j 2\pi \frac{k}{N_{\mathrm{c}}} n} e^{-j \pi \frac{l^2}{N_{\mathrm{p}}}}\\
	&=r[n]e^{-j \pi \frac{l^2}{N_{\mathrm{p}}}}.
\end{aligned}
\label{eq:subcarrier_and_chirp}
\end{equation}

This result reveals that any arbitrary $m$-th AFDM subcarrier is mathematically equivalent to the $0$-th subcarrier subjected to a cyclic delay $l$ and a Doppler-induced phase rotation $e^{-j 2\pi kn/N_{\mathrm{c}}}$ as shown in Fig. \ref{fig:pro_afdm_waveform_de} and Fig. \ref{fig:pro_afdm_waveform_do}, with an additional constant phase factor dependent only on the delay $e^{-j \pi l^2/N_{\mathrm{p}}}$. In essence, the entire set of AFDM subcarrier basis functions can be generated from a single fundamental periodic chirp by applying a grid of discrete delays and Doppler shifts, thereby mirroring a set of radar echoes.

\begin{remark}
AFDM subcarriers can be interpreted as a generalized and discretized form of up-chirp-based FMCW radar waveforms as \eqref{eq:sub0_and_chirp} and \eqref{eq:subcarrier_and_chirp}.
\end{remark} 

The subcarrier index $m$ directly encodes delay and Doppler information, which provides AFDM with an innate capability to resolve a channel in the DAFT domain. Physically, this implies that the AFDM modulation process maps symbols $x[m]$ from the DAFT domain onto time-domain chirp waveforms, each with distinct DD characteristics. A delay of $l$ and a Doppler index of $k$ in the channel correspond to cyclic shifts of $Kl$ and $k$ positions, respectively, in the DAFT domain.

\section{AFDM Parameterized with Delay and Doppler}
\label{sec:ddafdm}


\subsection{Discrete Periodic Ambiguity Function Analysis}
\label{subsec:DPAF}
The discrete periodic AF (DPAF) evaluates the correlation of a signal against its cyclically time-shifted and frequency-shifted versions, making it ideal for waveforms that employ a cyclic prefix \cite{liu2025cpofdm,kebo2007ambiguity}. Moreover, since the AF is based on DD parameters, it inherently links channel parameters with DD-parameterized AFDM subcarriers, simplifying the derivation.
The DPAF is defined for a periodic signal $\psi[n]$ of length $N_{\mathrm{c}}$ as:
\begin{equation}
\Lambda^{\psi}[l, k] = \sum_{n=0}^{N_{\mathrm{c}}-1} \psi[n] \psi^*[(n - l)_{N_{\mathrm{c}}}] 
e^{j 2\pi k n / N_{\mathrm{c}}},
\end{equation}
where $l$ is the delay index, and $k$ is the Doppler index.

Our analysis proceeds in two stages. First, we show that the DPAF of any subcarrier, in both its auto-DPAF (AAF) and cross-DPAF (CAF) forms, can be expressed in terms of the AAF of the $0$th subcarrier, denoted as $\Lambda^{\psi_{0}}$. This reveals that all subcarriers inherit the fundamental ambiguity properties of a single base waveform. Second, we derive the final value for $\Lambda^{\psi_{0}}$, proving its ideal sparse structure.

The foundation for this analysis is the representation of an arbitrary subcarrier, indexed by a DD pair $(l,k)$, using the $0$-th subcarrier from \eqref{eq:subcarrier_and_chirp}:
\begin{equation}
\psi_{(l, k)}[n] = \psi_0[(n - l)_{N_{\mathrm{c}}}] e^{-j 2 \pi \frac{n k}{N_{\mathrm{c}}}} e^{-j \pi \frac{l^2}{N_{\mathrm{p}}}}.
\label{eq:subcarrier_representation_main}
\end{equation}
By substituting this into the definitions of the AAF and CAF, we obtain the key relationships that govern the ambiguity structure of the entire signal set. The detailed derivations are provided in Appendix~\ref{app:dpaf_derivation}. The results are:
\begin{align}
\Lambda^{\psi_{(l_p, k_p)}}[l, k] &= e^{j 2 \pi \frac{k l_p - l k_p}{N_{\mathrm{c}}}} \Lambda^{\psi_0}[l, k], \label{eq:AAF_relation_main} \\
\begin{split}
\Lambda^{\psi_{(l_p,k_p)},\psi_{(l'_p,k'_p)}}[l, k] &= e^{j2\pi \left( \frac{kl_p - lk'_p}{N_{\mathrm{c}}} + \frac{(k'_p - k_p)l_p}{N_{\mathrm{c}}} + \frac{l'^2_p - l^2_p}{2N_{\mathrm{p}}} \right)}\\
&\quad \times \Lambda^{\psi_0}[l + l_p' - l_p, k + k_p' - k_p].
\end{split} \label{eq:CAF_relation_main}
\end{align}
These equations elegantly demonstrate that the ambiguity properties are entirely determined by $\Lambda^{\psi_0}$. The derivation for $\Lambda^{\psi_0}$ itself, detailed in Appendix~\ref{app:dpaf_derivation}, reveals a sparse, thumbtack-like profile. Cause $\Lambda^{\psi_0}\neq0$ only when $k \equiv (0)_K \text{ and } l \equiv -\left(\left\lfloor \frac{k}{K} \right\rfloor\right)_{N_p}$, substituting it to \eqref{eq:CAF_relation_main}, we have:
\begin{equation}
\Lambda^{\psi_{(l_p, k_p)}, \psi_{(l_p', k_p')}}[l, k] =
    \begin{cases} 
    N_{\mathrm{c}} e^{j \phi_1} e^{j \phi_2}, & \text{if conditions met} \\
    0, & \text{otherwise}
\end{cases}
\label{eq:final_CAF_main}
\end{equation}
where $\phi_1 = 2\pi \left( \frac{kl_p - lk'_p}{N_{\mathrm{c}}} + \frac{(k'_p - k_p)l_p}{N_{\mathrm{c}}} + \frac{l'^2_p - l^2_p}{2N_{\mathrm{p}}} \right)$, $\phi_2 = -\pi \frac{(l + l_p' - l_p)^2}{N_{\mathrm{p}}}$, and the non-zero conditions are $k + k_p' - k_p \equiv (0)_{K}$ and $l + l_p' - l_p \equiv -(\lfloor \frac{k + k_p' - k_p}{K} \rfloor)_{N_{\mathrm{p}}}$.

The sparse nature of this function enables high-resolution sensing. However, the structure also reveals an inherent coupling between delay and Doppler, manifested by the floor function $\lfloor \cdot \rfloor$ in the conditions. This term shows that a large Doppler shift can induce an apparent delay offset, a fundamental characteristic of chirp-based radar signals.
\subsection{Input-Output Relationship in the DD-DAFT Domain}
\label{subsec:IO_relation}
Unlike conventional TF representations, the proposed DD-DAFT framework explicitly maps DAFT indices to physical DD coordinates, unveiling the coupling behavior between time-varying channels and FMCW-based subcarriers.
Based on the DD parameterization, we derive a new expression for the AFDM input-output relationship. By mapping the linear subcarrier index $m$ to the DD index $(l, k)$ via $m = (-Kl-k)_{N_{\mathrm{c}}}$, the transmitted signal $s[n]$ becomes:
\begin{equation}
s[n]=\frac{1}{\sqrt{N_{c}}}\sum_{k=0}^{K-1}\sum_{l=0}^{N_{p}-1}X[l,k]\psi_{(l,k)}[n],
\label{eq:TX_signal_DD}
\end{equation}
where $X[l,k]$ are the data symbols in the DD-DAFT domain, which is a new description of the DAFT domain, mapping its one-dimensional DAFT index to a two-dimensional representation using DD. After passing through the multipath channel, the demodulated symbol $Y[l,k]$ is given by (see Appendix~\ref{app:IO_derivation} for the derivation):
\begin{equation}
Y[l,k]=\sum_{i=1}^{P}h_{i}\sum_{l^{\prime}=0}^{N_{p}-1}\sum_{k^{\prime}=0}^{K-1}X[l^{\prime},k^{\prime}]A_{(l^{\prime},k^{\prime}),(l,k)}[l_{i},k_{i}] + W[l,k],
\label{eq:demod_general}
\end{equation}
where $W[l,k]$ is the post-processing AWGN and $A_{(\cdot)}$ is a coefficient related to the DPAF. Leveraging the sparse property of the DPAF, this simplifies to the final compact input-output relationship:
\begin{equation}
\begin{aligned}
    &Y[l,k] \\
    &= \sum_{i=1}^{P}h_{i}X[(l-l_{i}+\Delta l)_{N_{p}},(k-k_{i})_{K}] \cdot e^{j \phi_{ch}} + W[l,k],
\end{aligned}
\label{eq:DD_DAFT_IO}
\end{equation}
where the delay-coupling term is $\Delta l = \lfloor\frac{k-k_{i}}{K}\rfloor$ and the channel-induced phase is $e^{j \phi_{ch}} = e^{j2\pi\frac{(k-k_{i})l_{i}}{N_{c}}} e^{j\pi\frac{(2l-l_{i})l_{i}}{N_{p}}}$. This equation can be expressed as a 2D convolution:
\begin{equation}
Y[l,k]=\sum_{l^{\prime}=0}^{N_{p}-1}\sum_{k^{\prime}=0}^{K-1}X[l^{\prime},k^{\prime}]h_{w}[l,k;l^{\prime},k^{\prime}] + W[l,k],
\label{eq:IO_convolution}
\end{equation}
where the effect of channel is encapsulated in the kernel $h_{w}[\cdot]$:
\begin{equation}
\begin{aligned}
    &h_{w}[l,k;l^{\prime},k^{\prime}]\\
    &=\sum_{i=1}^{P}h_{i} e^{j \phi_{i}} \delta[(k^{\prime}-(k-k_{i}))_{K}] \delta[(l^{\prime}-(l-l_{i}+\Delta l'_{i}))_{N_{p}}],
\end{aligned}
\label{eq:hw_kernel}
\end{equation}
with the phase $\phi_{i} = 2\pi(\frac{l_{i}k^{\prime}}{N_{c}}+\frac{l^{\prime}l_{i}}{N_{p}}+\frac{l_{i}^{2}}{2N_{p}})$ and coupling $\Delta l'_{i} = \lfloor\frac{k-(k^{\prime}+k_{i})}{K}\rfloor$.

\subsection{Discussion on Signal Generation and Comparison}
\label{subsec:discussion}

We can examine the generation of the proposed FMCW-based AFDM signal from two perspectives, which also facilitates a detailed comparison with OTFS. 

The first perspective is based on physical waveform construction. The time-domain signal is formed by the superposition of $N_{\mathrm{c}}$ orthogonal FMCW-based subcarriers, each modulated by a data symbol. This generation method applies universally to multicarrier waveforms \cite{sahinSurveyMulticarrierCommunications2014}. For two-dimensional modulation schemes such as OTFS, this framework provides a clear comparison by replacing the AFDM chirp waveform with a periodic delta pulse train. Through substitution of the subcarrier representation \eqref{eq:subcarrier_representation_main} and the periodic chirp construction \eqref{eq:periodic_chirp_construction} into the transmit signal expression \eqref{eq:TX_signal_DD}, the AFDM modulated signal takes the form:
\begin{equation}
\begin{aligned}
s[n] &= \frac{1}{\sqrt{N_{c}}}\sum_{k=0}^{K-1}\sum_{l=0}^{N_{p}-1}X[l,k]\psi_0[(n - l)_{N_{\mathrm{c}}}] e^{-j 2 \pi \frac{n k}{N_{\mathrm{c}}}} e^{-j \pi \frac{l^2}{N_{\mathrm{p}}}}\\
&= \frac{1}{\sqrt{N_{c}}}\sum_{k=0}^{K-1}\sum_{l=0}^{N_{p}-1}X[l,k]\sum_{q=0}^{K-1} \tilde{\psi}_0[(n - l-qN_{\mathrm{p}})_{N_{\mathrm{p}}}]\\
&\quad \times e^{-j 2 \pi \frac{n k}{N_{\mathrm{c}}}} e^{-j \pi \frac{l^2}{N_{\mathrm{p}}}},
\label{eq:AFDMvsOTFS}
\end{aligned}
\end{equation}
where the constituent subcarriers are continuous periodic chirps:
\begin{equation}
    \psi_{l,k}[n] = \sum_{q=0}^{K-1} \tilde{\psi}_0[(n - l - qN_{\mathrm{p}})_{N_{\mathrm{p}}}] e^{-j2\pi\frac{n}{N_{\mathrm{c}}}k}e^{-j \pi \frac{l^2}{N_{\mathrm{p}}}}. \label{eq:afdm_chirp_subcarrier}
\end{equation}
Recalling the OTFS modulation formula \cite{bondre2022zak,lampel2022otfs}, its subcarriers are periodic delta pulses \cite{hadani2017orthogonal,linDelayDopplerPlaneOrthogonal2022,li2023pulsea}:
\begin{equation}
\psi_{l,k}^{\mathrm{OTFS}}[n] = \sum_{q=0}^{K-1} \delta[n - l - qL]_{L} e^{j2\pi\frac{l + qL}{KL}k}.\label{eq:otfs_delta_subcarrier}
\end{equation}

\begin{remark}
In this view, AFDM is analogous to FMCW radar, which uses continuous chirp signals \eqref{eq:afdm_chirp_subcarrier}, while OTFS is analogous to traditional pulse-Doppler radar, which uses delta pulses \eqref{eq:otfs_delta_subcarrier}.
\end{remark}
This highlights a key difference that AFDM uses a waveform with a $100\%$ duty cycle within each chirp period, potentially reducing the requirements for power amplifiers and offering better resilience against interference compared to the impulsive nature of OTFS.

The second perspective is implementation via discrete transforms. Here, the time-domain waveform is obtained efficiently by processing the modulation vector through an IDAFT. With our proposed parameterization ($c_2=0$), this simplifies to performing a standard inverse fast Fourier transform (IFFT), followed by a point-wise multiplication with a single diagonal phase matrix determined by $c_1$. This view reveals AFDM as an FMCW-windowed variant of OFDM. This structure not only underscores the high compatibility of AFDM with contemporary OFDM systems but also positions it as a simplified implementation of OCDM in certain configurations.

For a detailed comparison, we recall the OTFS input-output relationship with ideal pulses \cite{raviteja2018interference}:
\begin{equation}
 y[k, l] = \sum_{i=1}^{P} h_{i} e^{-j 2 \pi k_i l_i} x\left[\left(k - {k_{i}}\right)_{K}, \left(l - {l_{i}}\right)_{L}\right]. 
\end{equation}
The input-output relationship of AFDM in the DD-DAFT domain \eqref{eq:DD_DAFT_IO} exhibits a structure strikingly similar to OTFS, namely, a two-dimensional circular convolution between the transmitted signal and the channel. However, AFDM introduces an additional Doppler-delay coupling term, and the phase component differs due to the distinct underlying waveform structures. This is much like how OTFS with rectangular pulses also incurs additional phase terms compared to the ideal pulses case \cite{raviteja2018interference}. 
The key distinction, beyond phase differences, is that AFDM needs to address the additional Doppler coupling term. Interestingly, this coupling relationship also offers a novel perspective for explaining why AFDM can achieve lower channel estimation overhead cause the delay and Doppler can be fully characterized in one delta function \eqref{eq:hw_kernel}.

\section{Estimation Method of Target Parameters for AFDM-ISAC System}
\label{sec:algorithms}

\subsection{TF Domain Sensing Algorithms}
\label{subsec:tf_algorithms}
In monostatic sensing scenarios, algorithms operating in the TF domain provide a robust and efficient framework. The primary approach is the TF domain MF (TFMF), which is a general method based on correlating the received signal with a reference copy of the transmitted signal to achieve pulse compression. This method is effective even when the transmitted signal is random (e.g., data-carrying), provided the receiver can acquire or reconstruct it.

The TFMF process begins by reshaping the one-dimensional received signal $r[n]$ and transmitted signal $s[n]$ into two-dimensional matrices $r[n, q]$ and $s[n, q]$ of size $N_{\mathrm{p}} \times K$, where $n$ is the fast-time index and $q$ is the slow-time index.
\begin{equation}
s[n, q] = s[n + q N_{\mathrm{p}}], \quad r[n, q] = r[n + q N_{\mathrm{p}}].
\end{equation}

Next, an $N_{\mathrm{p}}$-point FFT is performed along the fast-time dimension $n$ for each column of $s[n, q]$ and $r[n, q]$ to obtain their frequency-domain representations, $S_{\text{fre}}[m, q] = \frac{1}{\sqrt{N_{\mathrm{p}}}} \sum_{n=0}^{N_{\mathrm{p}}-1} s[n, q] e^{-j \frac{2\pi}{N_{\mathrm{p}}} m n}$ and $R_{\text{fre}}[m, q] = \frac{1}{\sqrt{N_{\mathrm{p}}}} \sum_{n=0}^{N_{\mathrm{p}}-1} r[n, q] e^{-j \frac{2\pi}{N_{\mathrm{p}}} m n}$, where $m$ represents the fast-TF index. A matched filtering operation is then executed $\text{MF}[m, q] = R_{\text{fre}}[m, q] \cdot S_{\text{fre}}^*[m, q]$. 
To recover the delay information, an $N_{\mathrm{p}}$-point IFFT is applied to each column of the result along the fast-time frequency dimension $m$, yielding the range-processed matrix $D_{\text{Range}}[l, q] = \frac{1}{\sqrt{N_{\mathrm{p}}}} \sum_{m=0}^{N_{\mathrm{p}}-1} \text{MF}[m, q] e^{j \frac{2\pi}{N_{\mathrm{p}}} m l}$, where the peaks in index $l$ correspond to target delays.

Finally, Doppler information is extracted by performing a $K$-point IFFT on each row of $D_{\text{Range}}[l, q]$ along the slow-time dimension $q$. This produces the final DD map (DDM):
\begin{equation}
Z_{\text{TFMF}}[l, k] = \frac{1}{\sqrt{K}} \sum_{q=0}^{K-1} D_{\text{Range}}[l, q] e^{j \frac{2\pi}{K} q k}.
\end{equation}

A noteworthy special case of the general TFMF occurs when the transmitted signal is a known, deterministic waveform, such as a designated sensing pilot. In this scenario, the TFMF procedure can be simplified into the well-known and computationally efficient dechirp algorithm. This method, derived from FMCW radar principles, replaces the frequency-domain MF (which involves two FFTs and one IFFT for range processing) with a single time-domain multiplication followed by one FFT. This is possible because multiplying the received signal by the conjugate of the deterministic transmitted chirp directly removes its quadratic phase modulation, converting echoes with different delays into single-frequency components.

When the $0$-th subcarrier is designated as a sensing pilot ($s_0[n]$), the dechirp operation involves computing the element-wise product of the reshaped received signal $r[n, q]$ and the conjugate of the known pilot $s^*[n, q]$ to obtain the dechirp signal $d[n, q] = r[n, q] \cdot s^*[n, q]$.

Subsequently, an $N_{\mathrm{p}}$-point FFT is applied to each column of $d[n, q]$ to perform range processing, followed by a $K$-point IFFT along each row for Doppler processing, ultimately yielding the DDM $Z_{\text{Dechirp}}[l, k]$. 
Both the general TFMF and its dechirp variant have a computational complexity of $\mathcal{O}(N_{\mathrm{c}} \log N_{\mathrm{c}})$. While efficient, their primary limitation is that they operate in the TF domain and thus fail to explicitly compensate for the inherent DD coupling effects of chirp-based waveforms.
\subsection{DD-DAFT Domain Sensing Processing Algorithm}
To overcome the limitations of traditional TF domain methods, which do not fully exploit the DAFT structure of AFDM signals and tend to ignore DD coupling effects, a novel algorithm is proposed that performs MF directly in the DD-DAFT domain. This approach is based on the precise DD-DAFT domain input-output relationship derived in \eqref{eq:DD_DAFT_IO}. It correlates the received DD-DAFT domain signal $Y[n, m]$ with the expected transmitted symbols $X[l, k]$ after accounting for all potential channel effects, including coupling and phase shifts, to optimally estimate the target parameters $(l, k)$.

For each hypothesized target position $(l, k)$, the MF output $Z_{\text{DDMF}}[l, k]$ is computed as follows:
\begin{equation}
\begin{aligned}
    &Z_{\text{DDMF}}[l, k] = \sum_{n=0}^{N_{\mathrm{p}}-1} \sum_{m=0}^{K-1} Y^*[n, m] \\
    &\times X\left[ \left(n - l + \left\lfloor \frac{m - k}{K} \right\rfloor\right)_{N_{\mathrm{p}}}, (m - k)_K \right] \phi_{\text{match}}(n, m, l, k),
\end{aligned}
\end{equation}
where $Y^*[n, m]$ is the conjugate of the received DD-DAFT symbol and $X[n, m]$ represents the known transmitted symbols. The indices account for the channel-induced shifts and coupling. The term $\phi_{\text{match}}(n, m, l, k)$ is a phase matching factor that ensures coherent integration and is defined as:
\begin{equation}
\phi_{\text{match}}(n, m, l, k) = \exp\left( j 2\pi \left( \frac{l(m - k)}{N_{\mathrm{c}}} + \frac{n l}{N_{\mathrm{p}}} - \frac{l^2}{2N_{\mathrm{p}}} \right) \right).
\end{equation}
Although this algorithm has a high computational complexity of $\mathcal{O}(N_{\mathrm{c}}^2)$ due to its nested loop structure, it provides high precision by operating directly in the DD-DAFT domain. This allows it to fully leverage the structure of the AFDM waveform and explicitly handle both coupling and phase information, making it highly suitable for complex channel environments or scenarios that require the detection of weak signals in the presence of strong interference.
\subsection{Complexity and Feasibility Discussion}
A critical aspect for practical implementation is the trade-off between the superior performance of the proposed DD-DAFT domain MF (DDMF) algorithm and its computational cost. The TF-domain methods, including the general TFMF and its dechirp variant, exhibit a complexity of $\mathcal{O}\left(N_{c}\log N_{c}\right)$ due to their reliance on FFT operations. Furthermore, the dechirp algorithm reduces the computational overhead of two FFTs and offers the advantage of intermediate frequency signal processing.

In stark contrast, the DDMF algorithm, with its exhaustive search structure, has a complexity of $\mathcal{O}\left(N_{c}^2\right)$.
This quadratic complexity presents a significant challenge for real-time processing, especially on resource-constrained devices. However, the DDMF algorithm serves two vital roles. First, it acts as a crucial theoretical benchmark, quantifying the optimal sensing performance achievable by fully exploiting the AFDM signal structure and providing a reference against which lower-complexity alternatives can be judged. Second, it is well-suited for applications where performance is paramount and computational resources are less constrained, such as in base station-centric sensing or offline data analysis. 


\section{Simulation Results}
\label{sec:simulation}

\begin{table}[htbp]
	\centering
	\caption{System Parameters for Simulation}
	\label{tab:sim_paramters}
	\begin{tabular}{l|c}
		\hline
		\textbf{Parameter}                                                      & \textbf{Values}                    \\ \hline
		Subcarrier number, $N_{\mathrm c}$                                                 & 512                                \\ \hline
		chirp number, $K$                                                         & 8                              \\ \hline
		chirp sample number, $N_{\mathrm p}$                                               &  64                            \\ \hline
		maximum delay tap, $l_{\mathrm{max}}$   										& 10 \\ \hline
		maximum Doppler tap, $k_{\mathrm{max}}$ 										& 3  \\ \hline
		AFDM parameter 1, $c_1$ 										& $\frac{K}{2N_{\mathrm{c}}}=\frac{1}{64}$, $\frac{2 k_{\mathrm{max}}+1}{2N_{\mathrm{c}}}=\frac{7}{1024}$  \\ \hline
		AFDM parameter 2, $c_2$ 										& $0$, $\sqrt{2}$  \\ \hline
	\end{tabular}
    \vspace{-10pt}
\end{table}

\begin{figure*}[htbp]
	\centering
	\subfloat[{TFMF with $\text{PO}=0$}]{\includegraphics[width=0.23\linewidth]{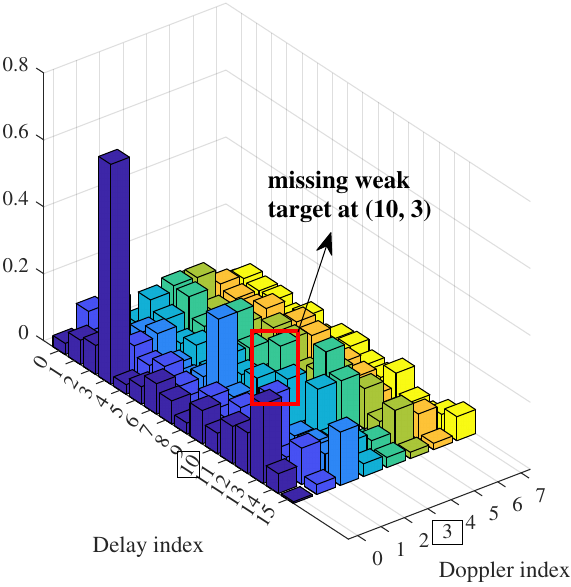}\label{fig:TFMF_overhead0}}
	\subfloat[{DDMF with $\text{PO}=0$}]{\includegraphics[width=0.23\linewidth]{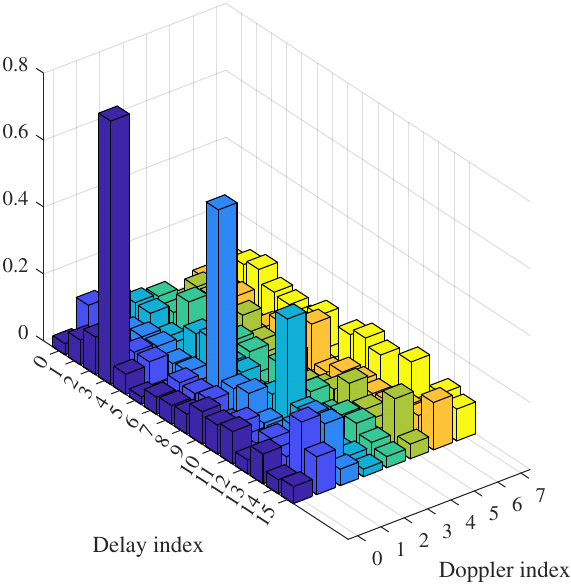}\label{fig:DDMF_overhead0}}
	\subfloat[{TFMF with $\text{PO}=1$}]{\includegraphics[width=0.25\linewidth]{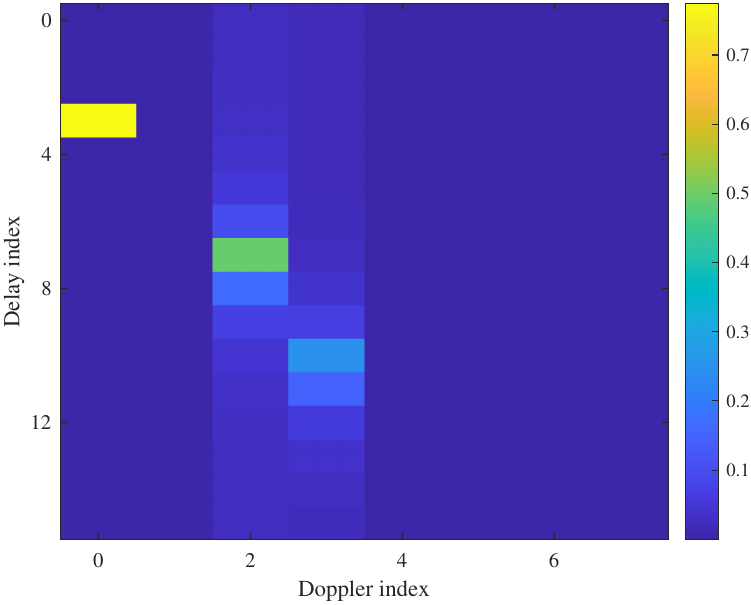}\label{fig:TFMF_overhead1}}
	\subfloat[{DDMF with $\text{PO}=1$}]{\includegraphics[width=0.25\linewidth]{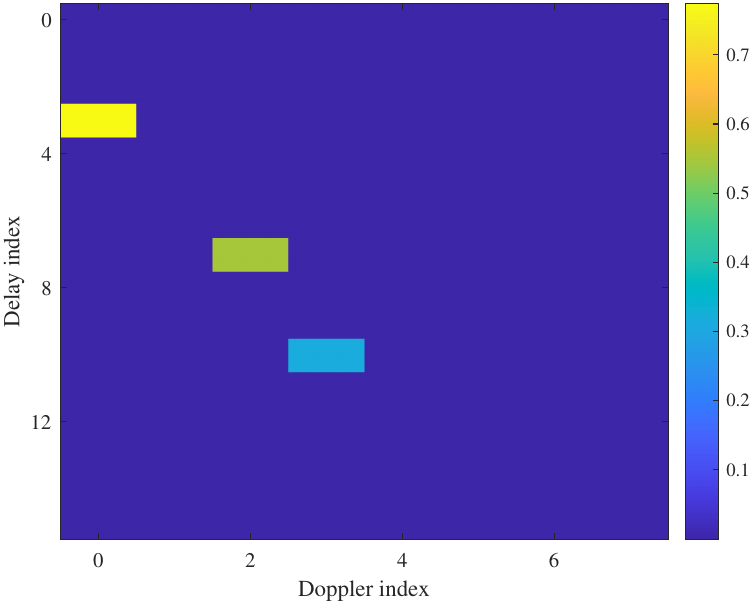}\label{fig:DDMF_overhead1}}
	\caption{DDMs of TFMF and DDMF algorithms for $\text{PO}=0$ and $\text{PO}=1$ cases.}
	\label{fig:MF_results}
    \vspace{-10pt}
\end{figure*}

In this section we consider an adjusted AFDM symbol with the proposed parameters $c_1=\frac{K}{2N_{\mathrm{c}}}=\frac{1}{2N_{\mathrm{p}}}$ and $c_2=0$, where $N_{\mathrm{c}}=KN_{\mathrm{p}}$. By contrast, the classic AFDM symbol uses $c_1=\frac{2k_{\mathrm{max}}+1}{2N_{\mathrm{c}}}$ and $c_2=\sqrt{2}$ as a baseline waveform \cite{bemani2023affineb} that has not been optimized for FMCW-based ISAC. Note that, apart from the fixed value of $c_2$, the proposed parameter-selection criterion chiefly determines the value of $c_1$, with $K$ chosen on the basis of the number of subcarriers $N_{\mathrm{c}}$ and the channel’s maximum Doppler tap $k_{\mathrm{max}}$, e.g., $K>2k_{\mathrm{max}},N_{\mathrm{p}}>l_{\mathrm{max}}$.

Information bits are generated randomly and mapped to 4QAM symbols.  The carrier frequency is set to $79$ GHz with $15$ kHz subcarrier spacing, yielding a Doppler resolution of $15$ kHz.
We consider AFDM with $N_{\mathrm{c}} = 512$ subcarriers, giving a bandwidth of $7.68$ MHz, which fixes the sample interval (delay resolution) at $130.2$ ns.
The maximum target range and velocity are taken to be $195.3$ m and $307.6$ km/h, corresponding to a round-trip delay of $1302$ ns and a Doppler shift of $45$ kHz.
This configuration highlights the hardware reducibility benefit of the proposed FMCW-based AFDM. Specifically, with the analog dechirping enabled by proposed $c_1,c_2$, the required ADC bandwidth is determined by the maximum beat frequency $B_{\text{eff}} \approx \mu \tau_{\max} + \nu_{\max}$. Substituting the simulation parameters yields $B_{\text{eff}} \approx 195$ kHz. Consequently, compared to the standard AFDM baseline which requires sampling the full 7.68 MHz bandwidth, our method reduces the ADC sampling rate requirement by a factor of approximately 39.4 ($7.68$ MHz / $0.195$ MHz), offering significant hardware efficiency.
According to the delay and Doppler sampling in \eqref{rangesample}, the maximum delay and Doppler taps are $l_{\mathrm{max}} = 10$ and $k_{\mathrm{max}} = 3$, respectively.  Unless otherwise specified, all simulation parameters are listed in Table \ref{tab:sim_paramters}.
Note that when a pilot is employed, fixed at the $0$-th subcarrier, the total energy of the AFDM symbol remains constant, reflecting a scenario with a power constraint. 
And the pilot overhead (PO) is defined as the proportion of subcarriers occupied by the pilot and the zero guard intervals $\text{PO} =\frac{2Q+\delta_{\mathrm{dc}}}{N_{\mathrm{c}}}$, where $Q$ is the number of single-sided zero subcarriers and $\delta_{\mathrm{dc}}\in\{0,1\}$ indicates the presence of a DC pilot. Specifically, when $\text{PO} = 1$, $2Q=N_{\mathrm{c}}-1$, and when $\text{PO} = 0$, data occupies all subcarriers. Note that TFMF and DDMF require all transmitted symbols $x[m]$ but are pilot-independent, whereas dechirp only requires the pilot $x[0]$. Additionally, all waveforms are generated via \eqref{eq:AFDM_modulation}. By using PO to describe the communication rate or spectral efficiency inherent in $x[m]$, a fair comparison is ensured across all schemes under equivalent TF resources and communication constraints.

We assume three targets with DD taps $(l,k)=[(3,0),(7,2),(10,3)]$ and powers $|h|^2=[0.6, 0.3, 0.1]$.
Fig. \ref{fig:TFMF_overhead0} and Fig. \ref{fig:DDMF_overhead0} depict the DDM for targets when transmitting an AFDM symbol without a pilot. It can be observed that both algorithms detect the strong targets, but the weak target located at $(10,3)$ is missed by TFMF.
In this case, data interference dominates the effectiveness of the algorithms. 
The activated data subcarriers prevent the chirp-based pilot signal from being identical from cycle to cycle, and the loss of phase coherence across the $K$ chirp periods degrades the processing gain of the Doppler FFT (the transform along the slow-time axis), causing spectral leakage that disperses the target energy in the DDM.
Fig. \ref{fig:TFMF_overhead1} and Fig. \ref{fig:DDMF_overhead1} present the DDM when only one pilot is transmitted. The results illustrate that the DD coupling inherent to the chirp waveform in the TF domain can be effectively handled by the DDMF algorithm.

\begin{figure*}[t]
	\centering
	\subfloat[{PSLR vs SNR}]{\includegraphics[width=0.33\linewidth]
{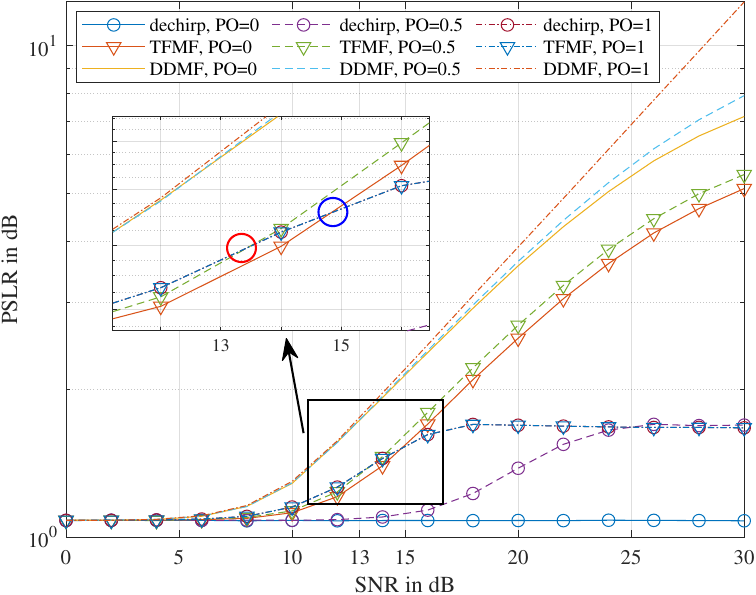}\label{fig:pslr_PO}}
	\subfloat[{Image SNR vs SNR}]{\includegraphics[width=0.33\linewidth]
{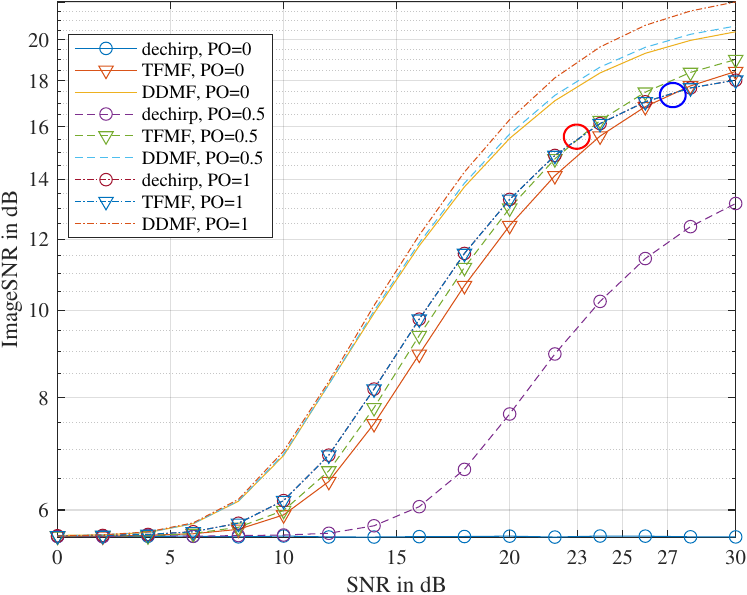}\label{fig:imagesnr_PO}}
	\subfloat[{Pd vs SNR}]{\includegraphics[width=0.33\linewidth]
{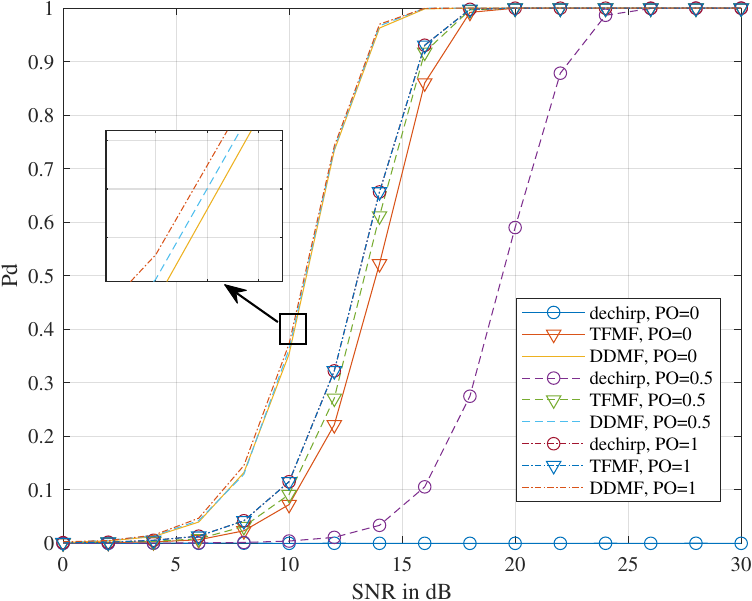}\label{fig:pd_PO}}
	\caption{PSLR, Image SNR, and Pd of the proposed AFDM under different PO and algorithms.}
	\label{fig:powerSNR_PO}
    \vspace{-10pt}
\end{figure*}

\begin{figure*}[htbp]
	\centering
    \subfloat[{Image SNR performance of the \\proposed AFDM and its variant}]{\includegraphics[width=0.248\linewidth]
{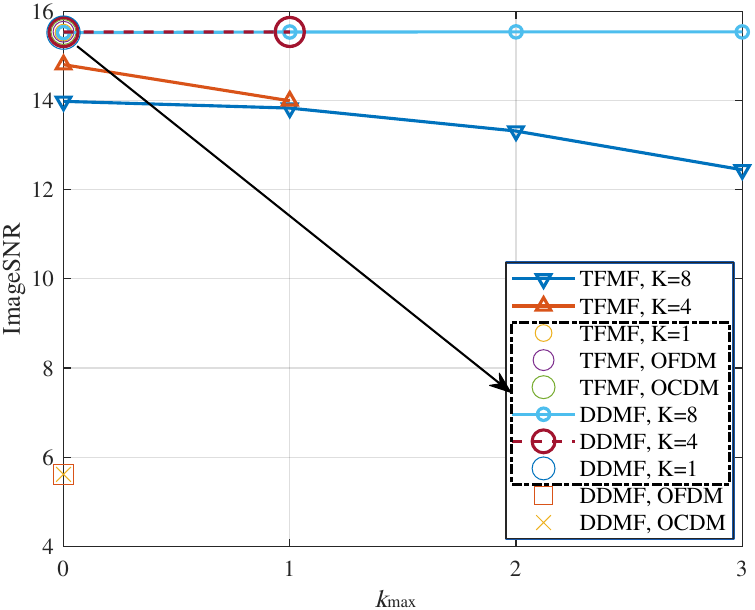}\label{fig:vsImageSNR}}
	\subfloat[{PSLR performance of the \\proposed AFDM and classic AFDM}]{\includegraphics[width=0.25\linewidth]
{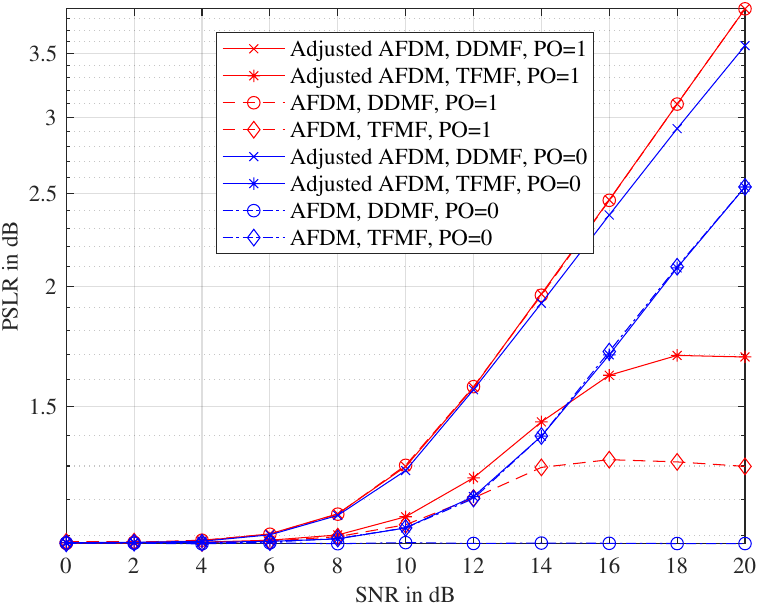}\label{fig:vspslr}}
	\subfloat[{Pd performance of the \\proposed AFDM and classic AFDM}]{\includegraphics[width=0.248\linewidth]
{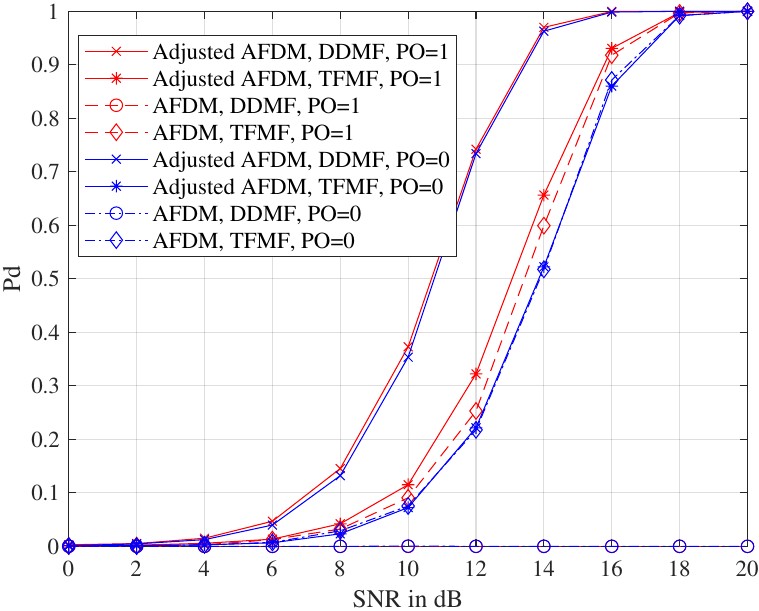}\label{fig:vspd}}
    \subfloat[{BER performance of the \\proposed AFDM and classic AFDM}]{\includegraphics[width=0.25\linewidth]
{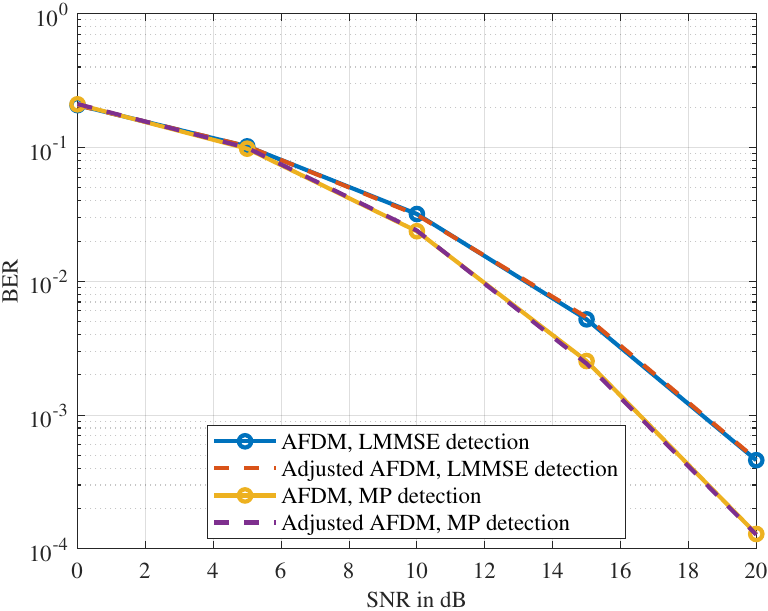}\label{fig:vsber_all}}
	\caption{The Image SNR, BER, PSLR, and Pd performance comparison of waveforms under TF and DD algorithms.}
	\label{fig:vs}
    \vspace{-10pt}
\end{figure*}

To evaluate the sensing performance using suitable metrics, we assume a single reflecting target located at a DD index of $(l,k)=(10,3)$ with a unit-power channel gain, i.e., $h = 1$. Fig. \ref{fig:pslr_PO} and Fig. \ref{fig:imagesnr_PO} show the peak-to-maximum sidelobe ratio (PSLR) and image SNR versus the transmitted symbol SNR, respectively. The image SNR represents the peak-to-average noise ratio with the radar processing gain. It can be observed that the DDMF outperforms the TFMF and the dechirp algorithm. The performance of each algorithm, with the exception of TFMF, improves as the PO increases. Meanwhile, the dechirp algorithm exhibits a performance floor as the SNR increases. The PSLR and image SNR curves for the different schemes show similar trends, as both metrics jointly characterize the degree of target energy-focusing in the DDM, but they respectively emphasize the target interference level and target detectability.

\begin{figure}[htbp]
	\centering
	\subfloat[{Pilot-based and pilot-free \\AFDM with the same useful rate}]{\includegraphics[width=0.5\linewidth]
{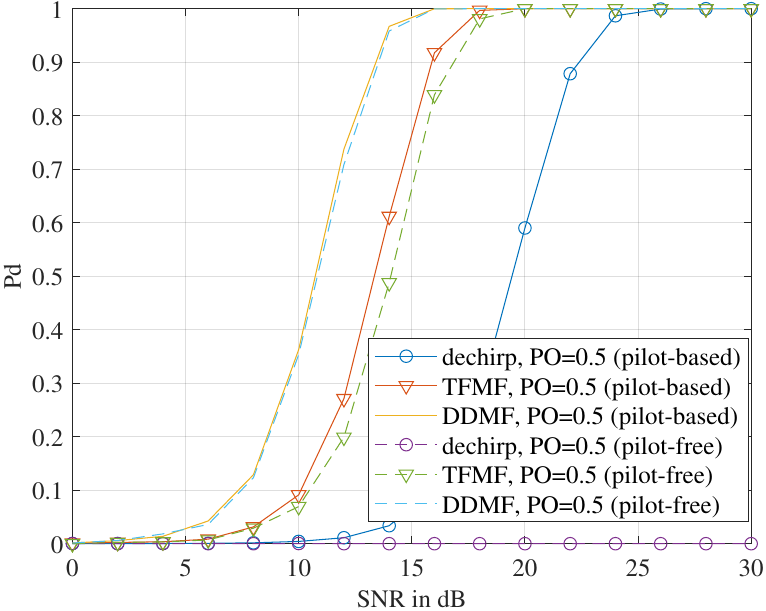}\label{fig:pilotfree}}
    \subfloat[{Proposed AFDM and classic AFDM in fractional DD channel}]{\includegraphics[width=0.5\linewidth]
{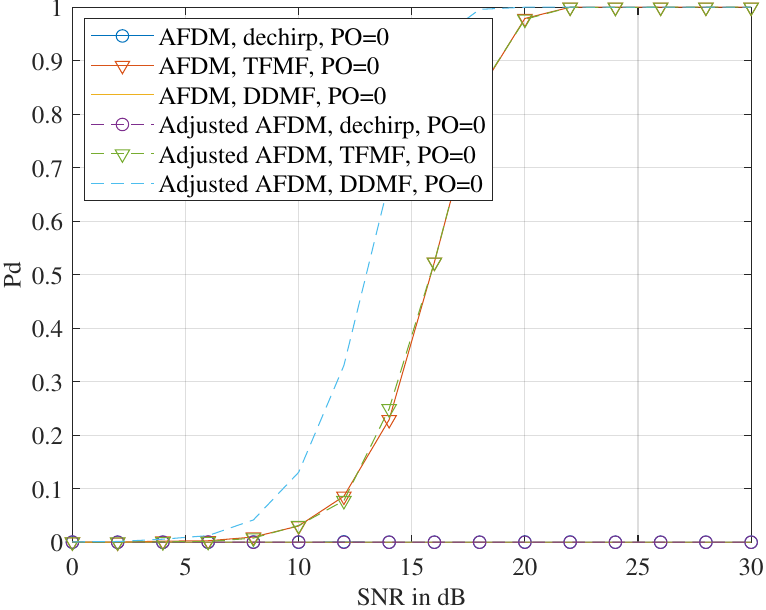}\label{fig:frac}}
	\caption{Pd performance comparison.}
	\label{fig:pdvs}
    \vspace{-10pt}
\end{figure}

The ability of DDMF to perfectly model the signal-channel interaction and exhaustively solve for the parameters allows it to achieve optimal performance, albeit at the cost of the highest computational complexity.
For the TFMF, its energy focusing performance does not monotonically increase with the PO. This is because data symbols introduce interference while simultaneously suppressing the DD coupling phenomenon. These two effects dominate at low and high SNRs, respectively, with the turning point corresponding to the intersection of the $\text{PO}=0$ and $\text{PO}=1$ curves (blue circle). As a trade-off, the $\text{PO}=0.5$ curve performs slightly worse than the $\text{PO}=1$ curve at low SNRs but significantly better at high SNRs (intersection at the red circle). This illustrates that signals with high data interference (low PO) perform worse at low SNRs, whereas signals with significant DD coupling (high PO) perform worse at high SNRs. Moreover, the $\text{PO}=0.5$ curve consistently outperforms the $\text{PO}=0$ curve, suggesting that its level of data interference is manageable and beneficially suppresses the DD coupling that the TFMF cannot handle, resulting in a lower net performance degradation.
For the dechirp algorithm, a low-complexity implementation of TFMF, its performance is identical to TFMF at $\text{PO}=1$ but degrades significantly as PO decreases. This is because the algorithm exclusively uses the deterministic pilot portion for matching and is entirely unable to process the interference from data symbols. Consequently, the benefit of DD coupling suppression from the data symbols cannot overcome the significant impact of their interference. In contrast, the TFMF can leverage data to obtain an accurate real-time delay estimate, with the data interference for TFMF primarily manifesting during phase matching and Doppler processing.

Fig. \ref{fig:pd_PO} illustrates the probability of detection (Pd) for the different schemes. For target detection, a 2D cell-averaging constant false alarm rate (CA-CFAR) detector [\citealp{richards2014radar}, Chapter 6.5.2] is employed, configured with $2$ training cells, $1$ guard cell, and a probability of false alarm (Pfa) of $10^{-4}$. The Pd curves show that all algorithms, including TFMF, consistently exhibit better performance as the PO increases. When compared with Fig. \ref{fig:imagesnr_PO}, this behavior is consistent with the trend at lower SNRs. The reason the trend for TFMF does not reverse at higher SNRs is that the target energy in the DDM becomes sufficiently high to achieve a Pd of $1$ before the SNR reaches the high-value region (approx. $23$ dB) where performance degradation would otherwise occur. This reveals that with an effective CFAR detector, increasing the proportion of data symbols (i.e., a lower PO) leads to a decrease in detection performance, highlighting a fundamental trade-off between communication rate and sensing performance.

%


In Fig. \ref{fig:pilotfree}, we compare the proposed AFDM at $\text{PO} = 0.5$ under both pilot-assisted and pilot-free conditions. The simulation results show that, with the exception of dechirp, effective sensing can be achieved without pilots. The sensing performance consistently degrades in the order of DDMF, TFMF, and dechirp. Furthermore, for the same algorithm, pilot-assisted schemes consistently outperform pilot-free ones, confirming that deterministic pilots provide a tangible gain for sensing. Besides, using the current system parameters, we evaluated a target located at a non-integer grid point $(l, k) = (10.2, 2.7)$. Fig. \ref{fig:frac} demonstrates that the proposed algorithm remains effective. However, there is a performance degradation of approximately 2 dB compared to Fig. \ref{fig:powerSNR_PO}. This is attributed to the energy leakage caused by the fractional shifts, which results in a reduction of the target's peak magnitude.

To observe the adaptability of waveforms to the channel, we consider the maximum Doppler shift of the channel $k_{\mathrm{max}}=[0,1,2,3]$. We set the parameter $K=[1,4,8]$ for the proposed AFDM, along with the OFDM waveform ($c_1=c_2=0$) and the OCDM waveform ($c_1=c_2=\frac{1}{2N_{\mathrm{c}}}$). 
Fig. \ref{fig:vsImageSNR} shows that for the DDMF algorithm, the performance of AFDM is neither affected by the maximum Doppler shift of the channel nor by the value of $K$. However, for the TFMF algorithm, an increase in maximum Doppler shift leads to a degradation in AFDM performance. This occurs because an increase in channel Doppler causes more subcarrier Doppler indices to exceed the period $K$, thereby exacerbating the DD coupling. Additionally, an increase in $K$ also causes the performance of AFDM under the TFMF algorithm to degrade. This degradation is due to inter-period interference caused by the activation of non-zero subcarriers (interpretable as data subcarrier interference when considering pilots). The increase in $K$ simultaneously reduces the number of points per period $N_{\mathrm{p}}$. While TFMF can perform range measurements matched to the number of fast-time samples $N_{\mathrm{p}}$ per period, it cannot effectively utilize the $K$ slow-time samples across different periods for pulse integration or Doppler phase processing. Therefore, when $k_{\mathrm{max}}=0$ and $K=1$, AFDM under TFMF achieves its optimal performance. We also observe that under these conditions, the OCDM and OFDM waveforms achieve the same optimal performance. However, OCDM and OFDM perform poorly under the DDMF algorithm, as the proposed DDMF algorithm relies on the phase matching introduced by the $c_2$ parameter. Besides, it can be noted that only when the choice of $K>2k_\mathrm{max}$ satisfies the requirement of AFDM for path separability, its corresponding waveform can accurately detect the target. The physical explanation for waveform speed measurement capability lies in the slow time sampling of FMCW. Since the total time is fixed, a larger $K$ value results in a smaller sampling interval. This means the maximum unambiguous velocity ($k_\mathrm{max}$) increases.

To validate how our proposed parameters selection impact on sensing performance, Fig. \ref{fig:vspslr} and Fig. \ref{fig:vspd} compare the PSLR and Pd of adjusted AFDM and classic AFDM. 
The results demonstrate that the proposed AFDM consistently outperforms the classic AFDM under identical PO and algorithmic conditions. For TFMF, this superiority arises because the chirp signal underlying the proposed AFDM exhibits perfect periodicity, thereby inheriting the favorable properties of FMCW, as illustrated in Fig. \ref{AFDM_waveform}. For DDMF, the advantage stems from the proposed AFDM’s ability to exploit the correspondence between FMCW echoes and channel DD parameters—characteristics that classic AFDM cannot fully capture.
Fig. \ref{fig:vsber_all} compares the bit error rate (BER) performance of the proposed AFDM scheme with that of classical AFDM. The delay and Doppler taps of the scatterers are set as $(l,k)=[(3,0),(7,2),(10,3)]$, with amplitudes $h=[0.6,0.3,0.1]$ following a Rayleigh distribution. As shown, the performance of the two schemes is identical when using either linear minimum mean square error (LMMSE) or message passing (MP) detectors. This is because, while the two approaches utilize different subcarrier subsets, both configurations satisfy the path separability requirement.

\section{Conclusion}
\label{sec:conclusion}
This paper established a fundamental equivalence between AFDM and Nyquist-sampled FMCW waveform through a novel parameter selection criterion. By reformulating the AFDM input-output relationship via DD parameters, we revealed the inherent DD coupling effect and characterized the channel interaction as a two-dimensional convolution. This waveform-centric perspective enabled the design of two matched-filtering algorithms, with the DD-DAFT domain approach demonstrating superior sensing performance by explicitly resolving this coupling. 
A key insight from our analysis clarifies the subcarriers of AFDM and OTFS as distinct radar paradigms, with AFDM employing periodic full-cycle chirps analogous to FMCW radar and OTFS utilizing periodic delta pulses comparable to pulse radar. Our findings demonstrate that although AFDM achieves performance parity with conventional waveforms (e.g., OFDM) in static channels, its full potential emerges in dynamic environments.

%
%
%

\appendices
\section{Derivation of DPAF}
\label{app:dpaf_derivation}

\subsection{AAF of $\psi_{(l_p, k_p)}[n]$ and its relationship to AAF of $\psi_0[n]$}

The subcarrier is expressed as:
\begin{equation}
\psi_{(l_p, k_p)}[n] = \psi_0[(n - l_p)_{N_{\mathrm{c}}}] e^{-j 2\pi n k_p / N_{\mathrm{c}}} 
e^{-j \pi l_p^2 / N_{\mathrm{p}}},
\end{equation}
where $\psi_0[n] = e^{j \pi n^2 / N_{\mathrm{p}}}$ is the base chirp signal and $N_{\mathrm{c}} = K N_{\mathrm{p}}$.
The AAF is given by:
\begin{equation}
\Lambda^{\psi_{(l_p, k_p)}}[l, k] = \sum_{n=0}^{N_{\mathrm{c}}-1} \psi_{(l_p, k_p)}[n] 
\psi^*_{(l_p, k_p)}[(n - l)_{N_{\mathrm{c}}}] e^{j 2\pi k n / N_{\mathrm{c}}}.
\end{equation}



For the waveform component, performing the variable substitution $n' = (n - l_p)_{N_{\mathrm{c}}}$ (where the cyclic shift preserves the summation) gives:
\begin{equation}
\sum_{n'=0}^{N_{\mathrm{c}}-1} \psi_0[n'] \psi_0^*[(n' - l)_{N_{\mathrm{c}}}] e^{j 2\pi k n' / N_{\mathrm{c}}} 
= \Lambda^{\psi_0}[l, k] e^{j 2\pi k l_p / N_{\mathrm{c}}}.
\end{equation}

After adjusting for the phase in the substitution, the final AAF is:
\begin{equation}
\Lambda^{\psi_{(l_p, k_p)}}[l, k] = e^{j 2\pi (k l_p - l k_p)/N_{\mathrm{c}}} 
\Lambda^{\psi_0}[l, k].\label{eq:AAF}
\end{equation}
\subsection{CAF for subcarriers $\psi_{(l_p, k_p)}[n]$ and $\psi_{(l'_p, k'_p)}[n]$ and its relationship to AAF of $\psi_0[n]$}

The CAF is defined as:
\begin{equation}
\begin{aligned}
    &\Lambda^{\psi_{(l_p, k_p)}, \psi_{(l'_p, k'_p)}}[l, k] \\
    &= \sum_{n=0}^{N_{\mathrm{c}}-1} 
    \psi_{(l_p, k_p)}[n] \psi^*_{(l'_p, k'_p)}[(n - l)_{N_{\mathrm{c}}}] e^{j 2\pi k n / N_{\mathrm{c}}}.
\end{aligned}
\end{equation}

Substituting $n' = (n - l_p)_{N_{\mathrm{c}}}$ and $\psi_{(l_p,k_p)}$ yields:
\begin{equation}
\begin{aligned}
    \sum_{n'=0}^{N_{\mathrm{c}}-1} \psi_0[n'] \psi_0^*[(n' - (l + l'_p - l_p))_{N_{\mathrm{c}}}] 
    e^{j 2\pi (k + k'_p - k_p) n' / N_{\mathrm{c}}} \\
    = \Lambda^{\psi_0}[l + l'_p - l_p, k + k'_p - k_p].
\end{aligned}
\end{equation}

Thus, the CAF is:
\begin{equation}
\begin{aligned}
    \Lambda^{\psi_{(l_p, k_p)}, \psi_{(l'_p, k'_p)}}[l, k]=&
    e^{j 2\pi \left( \frac{k l_p - l k'_p}{N_{\mathrm{c}}} + \frac{l'^2_p - l_p^2}{2 N_{\mathrm{p}}} 
    + \frac{(k'_p - k_p) l_p}{N_{\mathrm{c}}} \right)} \\
    &\times \Lambda^{\psi_0}[l + l'_p - l_p, k + k'_p - k_p].\label{eq:CAF}
\end{aligned}
\end{equation}
\subsection{AAF of $\Lambda^{\psi_0}[l, k]$}
To derive the closed-form expression for $\Lambda^{\psi_0}[l, k]$, we begin with its definition:
\begin{equation}
\Lambda^{\psi_0}[l, k] = \sum_{n=0}^{N_{\mathrm{c}}-1} \psi_0[n] \psi_0^*[(n - l)_{N_{\mathrm{c}}}] e^{j 2 \pi \frac{k n}{N_{\mathrm{c}}}}.
\end{equation}

We decompose the time index $n$ as $n = s + t N_{\mathrm{p}}$, where $s \in [0, N_{\mathrm{p}} - 1]$ represents the intra-period sample index and $t \in [0, K - 1]$ is the period index. Due to the periodic construction of $\psi_0[n]$ from the base chirp $\tilde{\psi}_0[n] = e^{j \pi n^2 / N_{\mathrm{p}}}$, the signal terms simplify:
\begin{align}
\psi_0[s + t N_{\mathrm{p}}] &= \tilde{\psi}_0[s], \\
\psi_0^*[(s - l + t N_{\mathrm{p}})_{N_{\mathrm{c}}}] &= \tilde{\psi}_0^*[(s - l)_{N_{\mathrm{p}}}].
\end{align}

The summation can then be rewritten as:
\begin{equation}
\Lambda^{\psi_0}[l, k] =  \sum_{t=0}^{K - 1} e^{j 2 \pi \frac{t k}{K}}  \left( \sum_{s=0}^{N_{\mathrm{p}} - 1} \tilde{\psi}_0[s] \tilde{\psi}_0^*[(s - l)_{N_{\mathrm{p}}}] e^{j 2 \pi \frac{s k}{N_{\mathrm{c}}}} \right).
\end{equation}

The first term, a summation over $t$, is a geometric series that is non-zero only when $k$ is an integer multiple of $K$. Let $k = pK$ for some integer $p$. Under this condition, the sum evaluates to $K$, otherwise, it is $0$.
Under the condition $k=pK$, the second term (the inner sum) becomes:
\begin{align}
&\sum_{s=0}^{N_{\mathrm{p}} - 1} e^{j \pi s^2 / N_{\mathrm{p}}} e^{-j \pi (s-l)^2 / N_{\mathrm{p}}} e^{j 2 \pi \frac{s p}{N_{\mathrm{p}}}} \nonumber \\
&= \sum_{s=0}^{N_{\mathrm{p}} - 1} e^{j \pi (s^2 - (s^2 - 2sl + l^2)) / N_{\mathrm{p}}} e^{j 2 \pi \frac{s p}{N_{\mathrm{p}}}} \nonumber \\
&= e^{-j \pi l^2 / N_{\mathrm{p}}} \sum_{s=0}^{N_{\mathrm{p}} - 1} e^{j 2 \pi s (l+p) / N_{\mathrm{p}}}.
\end{align}
This is also a geometric series, which evaluates to $N_{\mathrm{p}}$ only when $(l+p)$ is an integer multiple of $N_{\mathrm{p}}$, i.e., $l+p \equiv (0)_{N_{\mathrm{p}}}$.

Combining these conditions, $\Lambda^{\psi_0}[l, k]$ is non-zero only if $k \equiv (0)_{K}$ and $l \equiv -(p) _{N_{\mathrm{p}}}\equiv -(\lfloor k/K \rfloor) _{N_{\mathrm{p}}}$. When these conditions are met, the value of the AAF is:
\begin{equation}
\Lambda^{\psi_0}[l, k] = K \cdot N_{\mathrm{p}} \cdot e^{-j \pi l^2 / N_{\mathrm{p}}} = N_{\mathrm{c}} e^{-j \pi l^2 / N_{\mathrm{p}}}.
\end{equation}
Therefore, the complete expression for the base AAF exhibits a sparse, thumbtack-like structure:
\begin{equation}
\Lambda^{\psi_0}(l, k) =
\begin{cases}
N_{\mathrm{c}} \, e^{-j \pi \frac{l^2}{N_{\mathrm{p}}}}, & \text{if } k \equiv (0)_K \text{ and } l \equiv -(\left\lfloor \frac{k}{K} \right\rfloor)_{N_{\mathrm{p}}} \\
0, & \text{otherwise}.
\end{cases}
\end{equation}

Then, substituting this into \eqref{eq:AAF} and \eqref{eq:CAF} provides the non-zero conditions and corresponding phases.

\section{Derivation of the General Input-Output Relationship}
\label{app:IO_derivation}

The AFDM demodulation at the receiver is defined as:
\begin{equation}
Y[l,k] = \frac{1}{\sqrt{N_{\mathrm{c}}}} \sum_{n=0}^{N_{\mathrm{c}}-1} r[n] \psi_{(l,k)}^*[n],
\end{equation}
where $r[n]$ is the received time-domain signal, modeled as:
\begin{equation}
r[n] = \sum_{i=1}^{P} h_i s[(n-l_i)_{N_{\mathrm{c}}}] e^{-j2\pi \frac{k_i n}{N_{\mathrm{c}}}} + w[n].
\end{equation}
Substituting the expression for $r[n]$ into the demodulation formula yields:
\begin{equation}
\begin{aligned}
    Y[l,k] =& \frac{1}{\sqrt{N_{\mathrm{c}}}} \sum_{n=0}^{N_{\mathrm{c}}-1} \left( \sum_{i=1}^{P} h_i s[(n-l_i)_{N_{\mathrm{c}}}] e^{-j2\pi \frac{k_i n}{N_{\mathrm{c}}}} + w[n] \right) \\
    &\times \psi_{(l,k)}^*[n].
\end{aligned}
\end{equation}
We can separate the signal and noise components:
\begin{equation}
\begin{aligned}
Y[l,k] =& \sum_{i=1}^{P} h_i \left( \frac{1}{\sqrt{N_{\mathrm{c}}}} \sum_{n=0}^{N_{\mathrm{c}}-1} s[(n-l_i)_{N_{\mathrm{c}}}] e^{-j2\pi \frac{k_i n}{N_{\mathrm{c}}}} \psi_{(l,k)}^*[n] \right) \\
&+ W[l,k],
\end{aligned}
\end{equation}
where the post-processed noise term is $W[l,k] = \frac{1}{\sqrt{N_{\mathrm{c}}}} \sum_{n=0}^{N_{\mathrm{c}}-1} w[n] \psi_{(l,k)}^*[n]$.
Next, we substitute the definition of the transmitted signal $s[n]$ from \eqref{eq:TX_signal_DD}:
\begin{equation}
s[(n-l_i)_{N_{\mathrm{c}}}] = \frac{1}{\sqrt{N_{\mathrm{c}}}} \sum_{k'=0}^{K-1} \sum_{l'=0}^{N_{\mathrm{p}}-1} X[l', k'] \psi_{(l',k')}[(n-l_i)_{N_{\mathrm{c}}}].
\end{equation}

Plugging this into the signal component of $Y[l,k]$ and rearranging the order of the summations, we get:
\begin{equation}
\begin{aligned}
&Y[l,k] = \sum_{i=1}^{P} h_i \sum_{k'=0}^{K-1} \sum_{l'=0}^{N_{\mathrm{p}}-1} X[l', k'] \\
&\times \left( \frac{1}{N_{\mathrm{c}}} \sum_{n=0}^{N_{\mathrm{c}}-1} \psi_{(l',k')}[(n-l_i)_{N_{\mathrm{c}}}] e^{-j2\pi \frac{k_i n}{N_{\mathrm{c}}}} \psi_{(l,k)}^*[n] \right) + W[l,k].
\end{aligned}
\end{equation}

The term in the large parentheses is the interaction coefficient $A_{(l',k'),(l,k)}[l_i,k_i]$, which equals to $\frac{1}{N_{\mathrm{c}}}(\Lambda^{\psi_{(l, k)}, \psi_{(l', k')}}[l_i, k_i])^*$. This leads to the final expression for the general input-output relationship:
\begin{equation}
Y[l,k]=\sum_{i=1}^{P}h_{i}\sum_{l^{\prime}=0}^{N_{p}-1}\sum_{k^{\prime}=0}^{K-1}X[l^{\prime},k^{\prime}]A_{(l^{\prime},k^{\prime}),(l,k)}[l_{i},k_{i}] + W[l,k],
\end{equation}
which is precisely \eqref{eq:demod_general}.

\bibliographystyle{IEEEtran}


 

\vspace{-3em}
\begin{IEEEbiography}[{\includegraphics[width=1in,height=1.25in,clip,keepaspectratio]{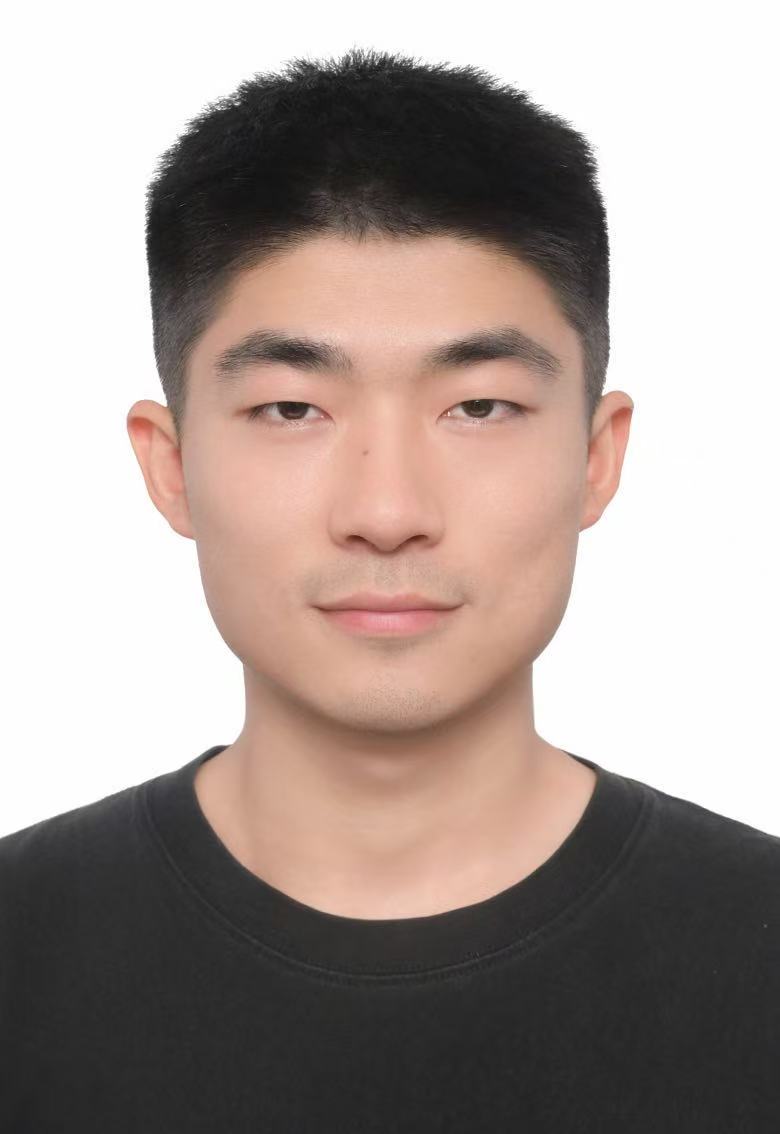}}]{Jiajun Zhu}
received the B.S. degree from the College of Information Science and Engineering, Hohai University, in 2022, and the M.S. degree from the School of
Electronics and Communication Engineering, Sun
Yat-sen University, China, in 2025. His research and engineering interests focus on digital integrated circuit design, waveform design, and integrated sensing and communications (ISAC).
\end{IEEEbiography}
\vspace{-3em}

\begin{IEEEbiography}[{\includegraphics[width=1in,height=1.25in,clip,keepaspectratio]{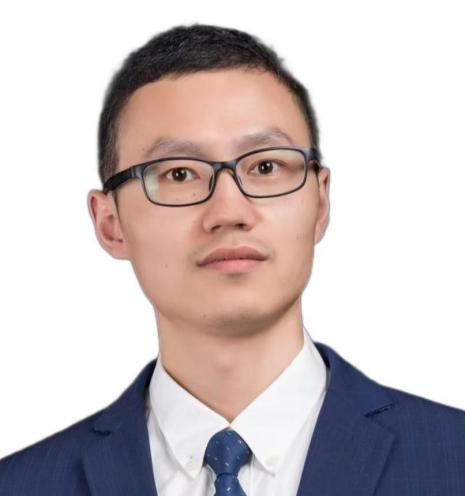}}]{Yanqun Tang}
(Member, IEEE) received his B.Sc., M.Sc., and Ph.D. degrees from the School of Electronic Science and Engineering, National University of Defense Technology, Changsha, China, in 2007, 2009, and 2013, respectively. He is currently an associate professor at the School of Electronics and Communication Engineering, Sun Yat-sen University, Shenzhen, China. He was a recipient of the IEEE PIMRC 2025 Best Paper Award and served as the technical program co-chair for the AFDM workshops in IEEE/CIC ICCC 2025. His research interests are integrated sensing and communication, full duplex communications, wireless physical-layer security, and machine learning techniques for wireless communications.
\end{IEEEbiography}
\vspace{-3em}

\begin{IEEEbiography}[{\includegraphics[width=1in,height=1.25in,clip,keepaspectratio]{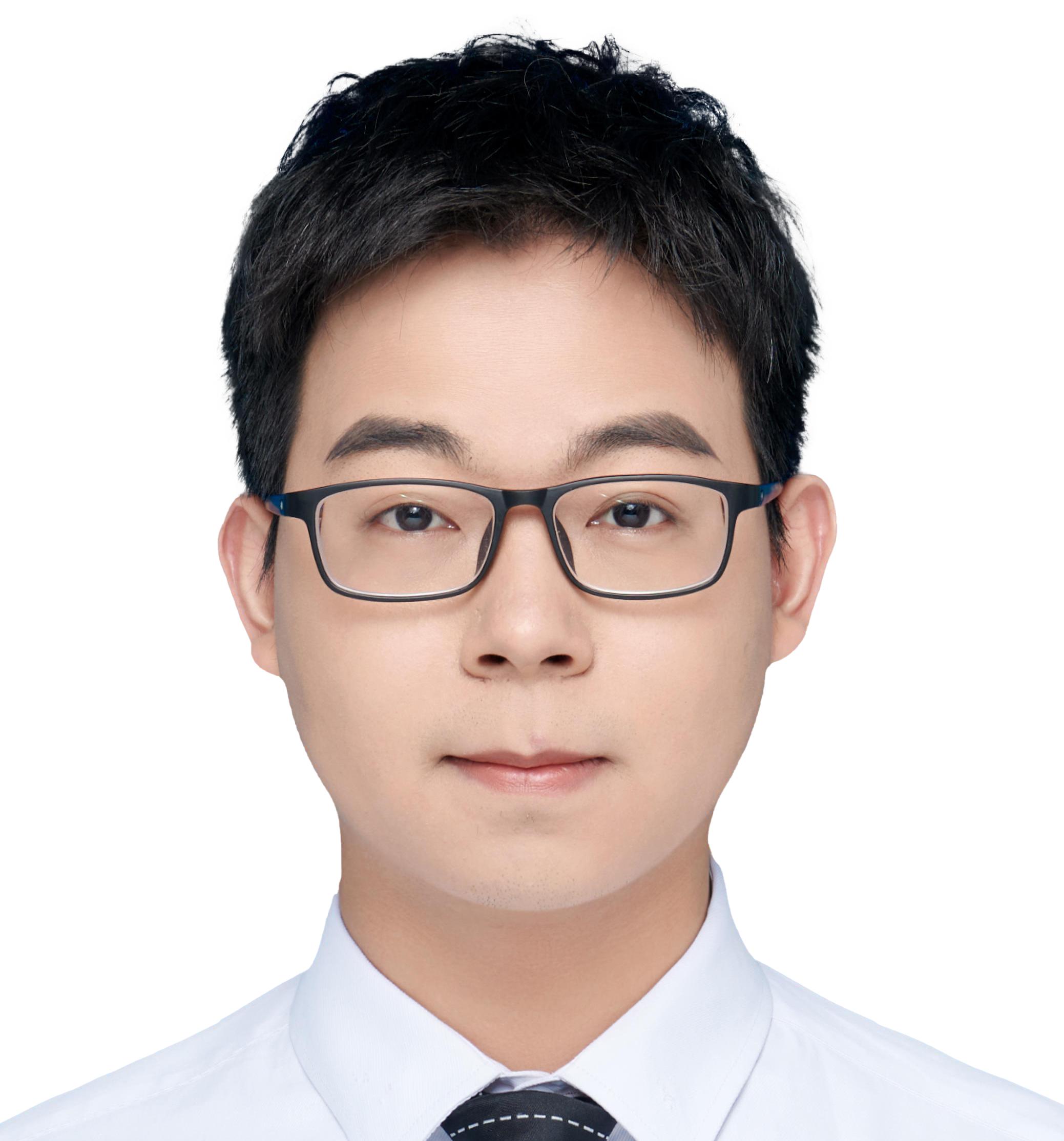}}]{Cong Yi}
received the M.S. degree in Electronic and Information Engineering from Sun Yat-sen University, Shenzhen, China, in 2025. He is currently pursuing the Ph.D. degree in Electronic and Information Engineering at Sun Yat-sen University. His research interests include advanced waveform design for next-generation wireless communication networks and integrated sensing and communications.
\end{IEEEbiography}

\begin{IEEEbiography}[{\includegraphics[width=1in,height=1.25in,clip,keepaspectratio]{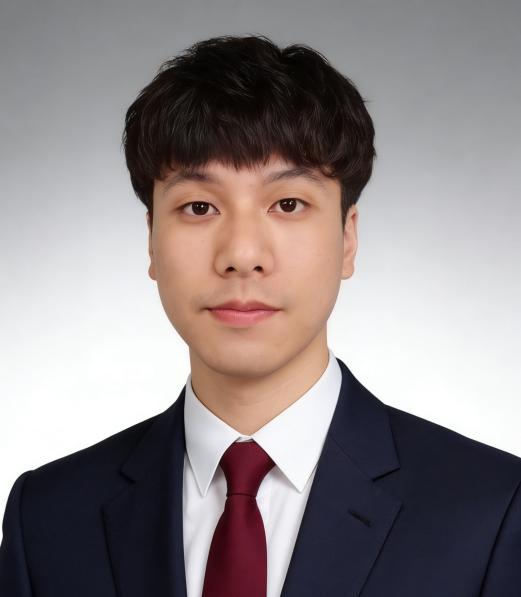}}]{Haoran Yin}
(Member, IEEE) received his B.Eng. and D.Eng. (Hons.) degrees from the School of Electronics and Communication Engineering, Sun Yat-Sen University, Shenzhen, China, in 2021 and 2026, respectively. From July 2025 to March 2026, he was a visiting Ph.D. student at the Wireless Research Lab, New York University (NYU) Abu Dhabi, Abu Dhabi, United Arab Emirates. 
He was a recipient of the IEEE PIMRC 2025 Best Paper Award and was awarded the China National Scholarship for two consecutive years. 
His research interests include advanced waveform design for Affine Frequency Division Multiplexing (AFDM) and Integrated Sensing and Communications (ISAC) for next-generation wireless communication networks.
\end{IEEEbiography}
\vspace{-3em}

\begin{IEEEbiography}[{\includegraphics[width=1in,height=1.25in,clip,keepaspectratio]{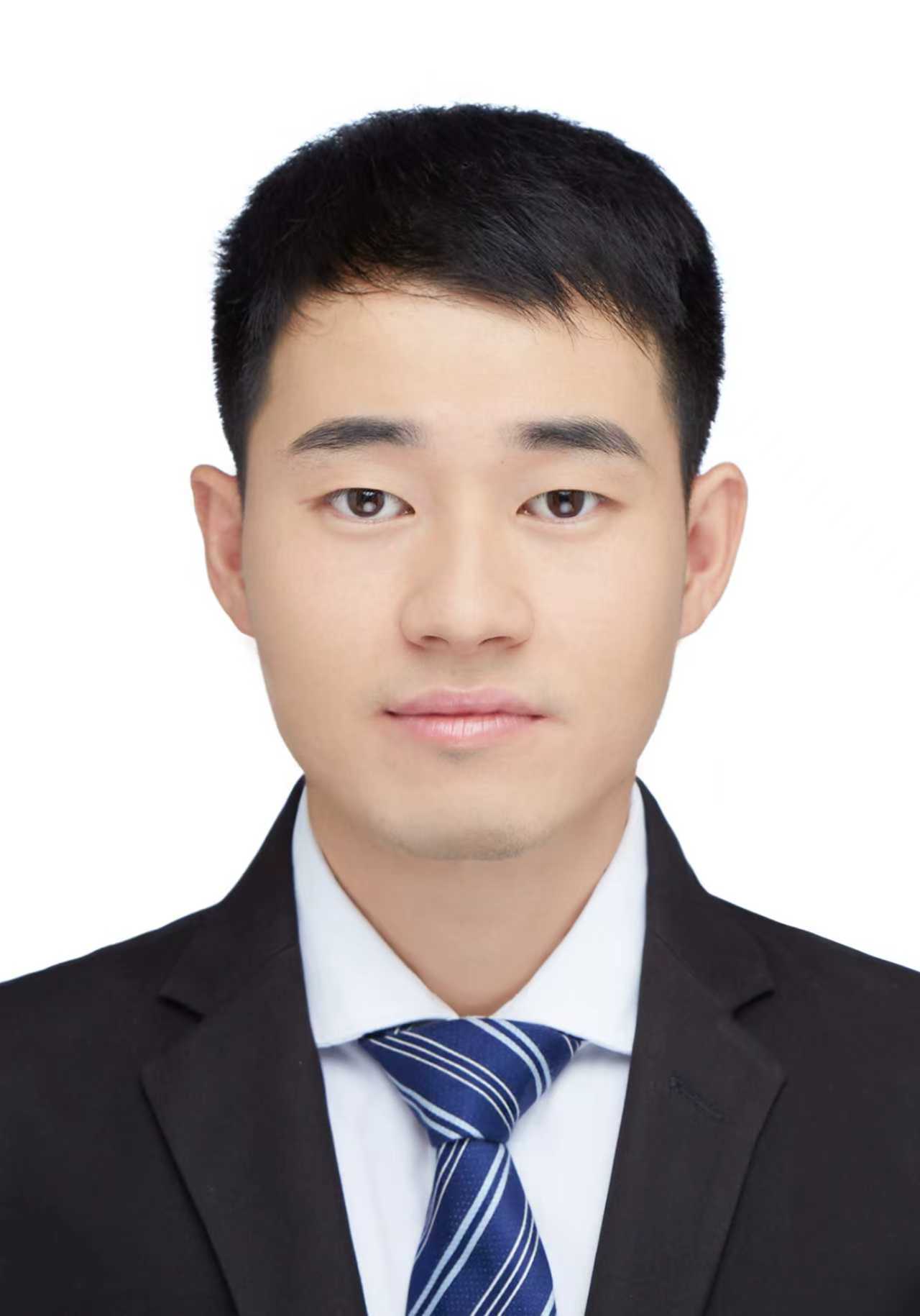}}]{Yuanhan Ni}
(Member, IEEE) received the B.E and Ph.D. degrees in electrical engineering from Beihang University (BUAA), Beijing, China, in 2017 and 2023, respectively. He is currently a Research Fellow with the School of Electronic and Information Engineering, Beihang University, Beijing, China. His research interests include Integrated Sensing and Communications (ISAC), physical layer security, affine frequency division multiplexing (AFDM) waveform. 
\end{IEEEbiography}
\vspace{-3em}

\begin{IEEEbiography}[{\includegraphics[width=1in,height=1.25in,clip,keepaspectratio]{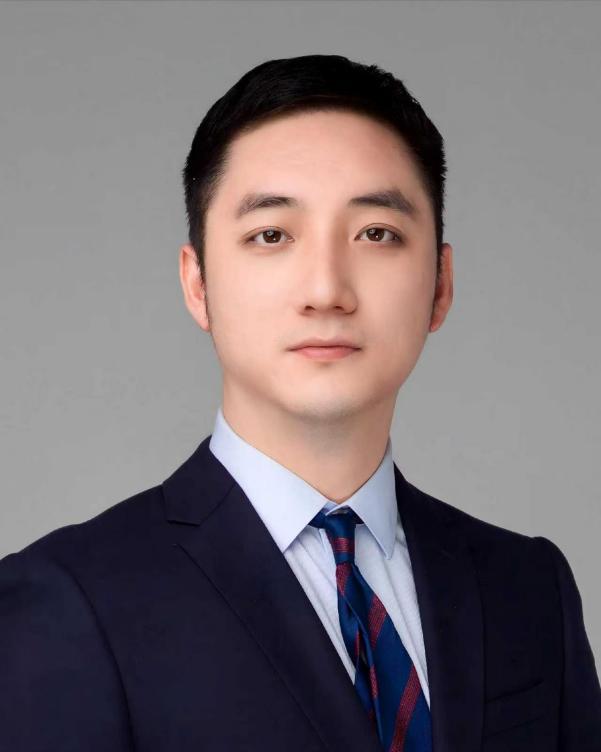}}]{Fan Liu}
(Senior Member, IEEE) is currently a full Professor with the National Mobile Communications Research Laboratory, School of Information Science and Engineering, Southeast University, Nanjing, China. Prior to that, he was an Assistant Professor with the Southern University of Science and Technology, Shenzhen, China, from 2020 to 2024. He received the Ph.D. and the BEng. degrees from Beijing Institute of Technology (BIT), Beijing, China, in 2018 and 2013, respectively. He has previously held academic positions in the University College London (UCL), London, UK, as a Visiting Researcher from 2016 to 2018, and a Marie Curie Research Fellow from 2018 to 2020. Prof. Liu's research interests lie in the general area of signal processing and wireless communications, and in particular in the area of Integrated Sensing and Communications (ISAC). 
\end{IEEEbiography}
\vspace{-3em}

\begin{IEEEbiography}[{\includegraphics[width=1.1in,height=1.25in,clip,keepaspectratio]{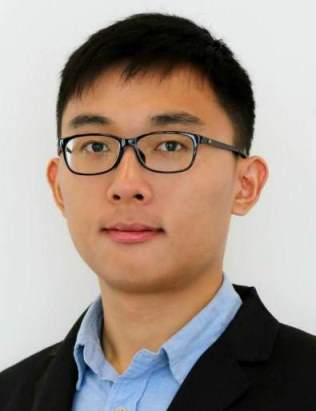}}]{Zhiqiang Wei} received the B.E. degree in information engineering from Northwestern Polytechnical University (NPU), Xi'an, China, in 2012, and the Ph.D. degree in electrical engineering and telecommunications from the University of New South Wales (UNSW), Sydney, Australia, in 2019. From 2019 to 2020, he was a Postdoctoral Research Fellow with UNSW. From 2021 to 2022, he was a Humboldt Postdoctoral Research Fellow with the Institute for Digital Communications, Friedrich-Alexander University Erlangen-Nuremberg (FAU), Erlangen, Germany. He is currently a Professor with the School of Mathematics and Statistics, Xi'an Jiaotong University, Xi'an, China. His research interests include delay-Doppler communications, resource allocation optimization, and statistical and array signal processing.  
\end{IEEEbiography}
\vspace{-3em}

\begin{IEEEbiography}[{\includegraphics[width=1in,height=1.25in,clip,keepaspectratio]{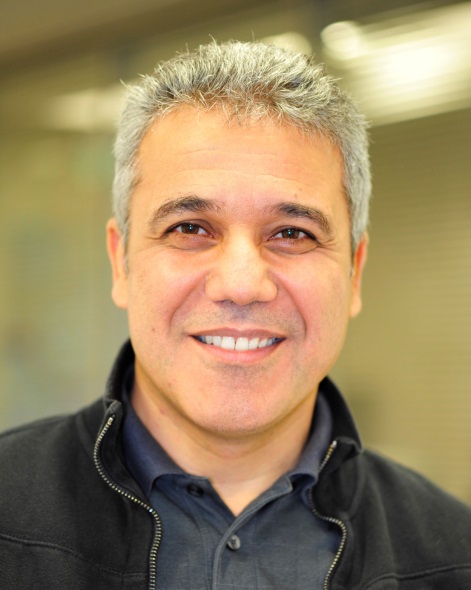}}]{H{\"u}seyin Arslan}
(Fellow, IEEE) received the B.S. degree in electrical and electronics engineering from Middle East Technical University, Ankara, T{\"u}rkiye in 1992, and the M.S. and Ph.D. degrees in electrical engineering from Southern Methodist University, Dallas, TX, USA, in 1994 and 1998, respectively. He is a Professor of Electrical Engineering and the Dean of the School of Engineering and Natural Sciences, Istanbul Medipol University, Istanbul, T{\"u}rkiye.
%
Dr. Arslan conducts research in wireless systems, with emphasis on the physical and medium access layers of communications. His current research interests are on 6G and beyond radio access technologies, physical layer security, interference management (avoidance, awareness, and cancellation), cognitive radio, multi-carrier wireless technologies (beyond OFDM), dynamic spectrum access, co-existence issues, non-terrestial communications (High Altitude Platforms), joint radar (sensing) and communication designs. 
%
\end{IEEEbiography}



\vfill

\end{document}